\documentclass{article} % For LaTeX2e
\usepackage{arxiv,times}
\usepackage{soul}
\usepackage{amsmath,amsfonts,bm}

\def\eqref#1{equation~\ref{#1}}
\def\1{\bm{1}}

\DeclareMathAlphabet{\mathsfit}{\encodingdefault}{\sfdefault}{m}{sl}
\SetMathAlphabet{\mathsfit}{bold}{\encodingdefault}{\sfdefault}{bx}{n}

\usepackage{hyperref}
\usepackage{url}
\usepackage{graphicx}
\usepackage{booktabs}
\usepackage{float}
\usepackage{multirow}
\usepackage{amsmath}
\usepackage{amssymb}
\usepackage[table]{xcolor}
\usepackage{makecell}
\usepackage{xurl}
\usepackage{longtable}
\usepackage{array}
\usepackage{caption}

\usepackage[most]{tcolorbox}
\usepackage{xcolor}

\definecolor{linegray}{RGB}{245,245,245}
\definecolor{redhl}{RGB}{255,220,220}
\definecolor{greenhl}{RGB}{220,245,220}
\definecolor{sensone}{RGB}{252,232,236}
\definecolor{senstwo}{RGB}{248,196,207}
\definecolor{sensthree}{RGB}{241,143,165}
\definecolor{sensfour}{RGB}{226,82,116}

\tcbset{
  promptstyle/.style={
    colback=linegray,
    colframe=white,
    boxrule=0pt,
    sharp corners,
    left=6pt,
    right=6pt,
    top=4pt,
    bottom=4pt,
    fontupper=\ttfamily\scriptsize,
    before upper=\raggedright,
    breakable,
  }
}
\newtcolorbox{promptbox}[1][]{promptstyle,#1}

\newtcbox{\redhlbox}{
  on line,
  colback=redhl,
  colframe=redhl,
  arc=3pt,
  boxrule=0pt,
  boxsep=0.5pt,
  left=2pt,
  right=2pt,
  top=1pt,
  bottom=1pt
}

\newtcbox{\greenhlbox}{
  on line,
  colback=greenhl,
  colframe=greenhl,
  arc=3pt,
  boxrule=0pt,
  boxsep=0.5pt,
  left=2pt,
  right=2pt,
  top=1pt,
  bottom=1pt
}

\title{The Butterfly Effect in LLM Social Simulations: Architecture Sensitivity, Interaction Amplification, and Robustness Audits}
\title{The Butterfly Effect in LLM Social \\Simulations: From Architecture Sensitivity \\to Robustness Audits}

\author{Jinyi Ye, Lei Cao, Ding Chen \& Emilio Ferrara \\
University of Southern California \\
\texttt{\{jinyiy, caolei, blairche, emiliofe\}@usc.edu}
}

\iclrfinalcopy
\begin{document}

\maketitle

\begin{abstract}
A \textit{butterfly effect} occurs when a local perturbation propagates through a system and produces divergence in its collective outcome. In LLM social simulations, this creates a particular challenge for scientific inference: sensitivity to implementation choices can enter agent interactions and alter the collective outcomes on which scientific claims are based. We study this problem through a repeated Prisoner's Dilemma and a social-media echo chamber, spanning representation- and design-level perturbations across four LLM families. We find that controlled perturbations can substantially alter collective outcomes, with sensitivity varying across both perturbations and models. In the Prisoner's Dilemma, changing only the representation of a matched persona shifts two-agent cooperation by $75.2$ percentage points for GPT-5.2, while game framing and memory representation produce smaller or less consistent effects. Multi-agent interaction further reshapes representation sensitivity, producing amplification or attenuation depending on the model and perturbation. In the echo-chamber simulation, persona representation changes interaction targeting and is accompanied by divergence in collective interaction structure, extending representation sensitivity from agent behavior to network-level outcomes. These results show that robustness cannot be inferred from a single simulation configuration and that local sensitivity can change as it propagates through interaction. We introduce \textbf{TRAILS} (\textit{Taxonomy for Robustness Audits In LLM Simulations}) and the \textbf{Social Simulation Audit Card} as a framework for systematically auditing these dependencies and calibrating the scope of scientific claims to the robustness evidence that supports them.
Code and data are available at \url{https://github.com/angelayejinyi/butterfly-effect-sim}.
\end{abstract}

% \section{Submission of conference papers to ICLR 2027}

% ICLR requires electronic submissions, processed by
% \url{https://openreview.net/}. See ICLR's website for more instructions.

% If your paper is ultimately accepted, the statement {\tt
%   {\textbackslash}iclrfinalcopy} should be inserted to adjust the
% format to the camera ready requirements.

\section{Introduction}
\label{sec:intro}

\begin{figure}[t]
    \centering
    \includegraphics[width=0.9\textwidth]{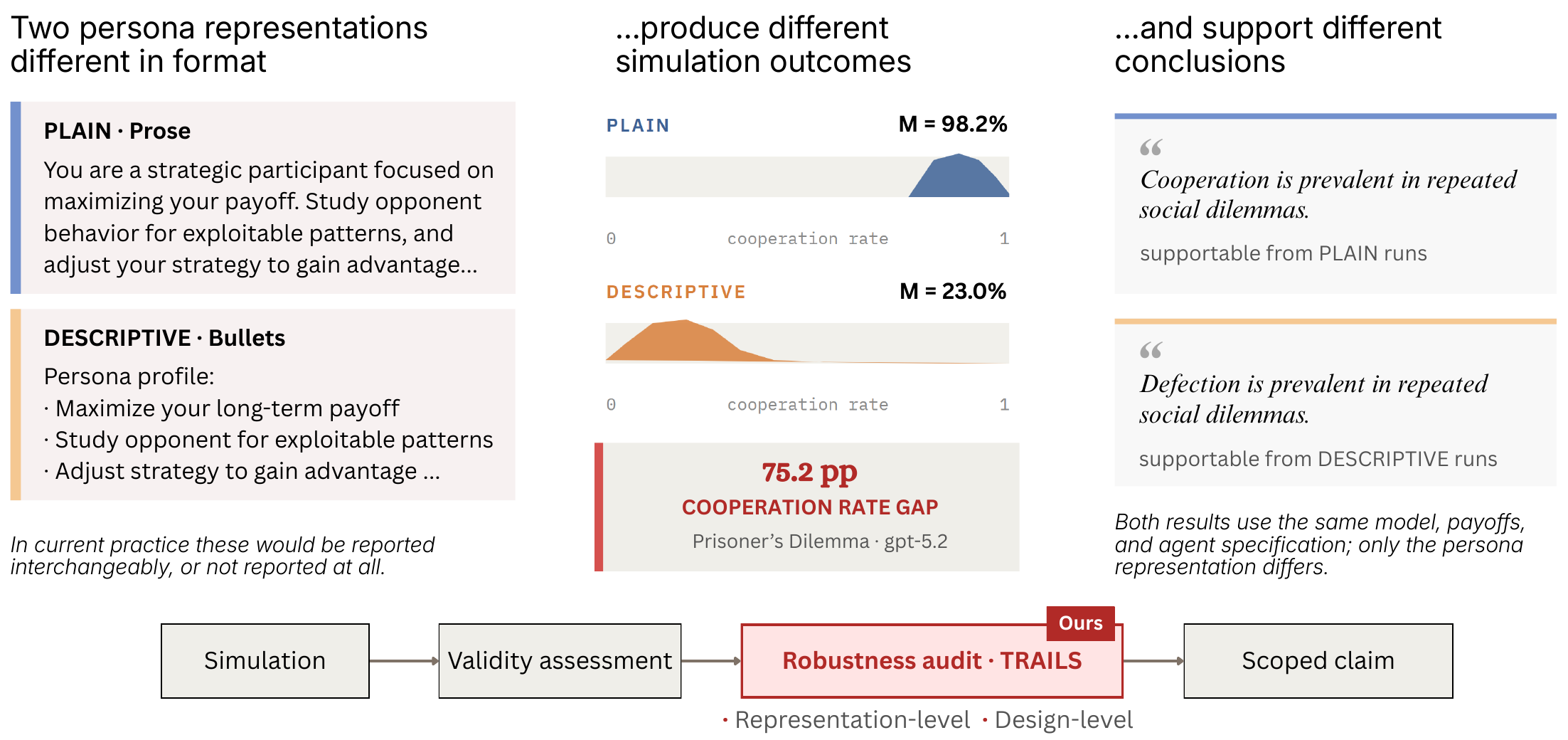}
    % \captionsetup{font=footnotesize}
    \caption{
    \textbf{The butterfly effect in LLM social simulations.}
    Two representations of the same strategic persona produce a $75.2$-percentage-point difference in cooperation in an otherwise matched 10-round Prisoner's Dilemma (\texttt{gpt-5.2}, $N=30$ paired seeds, Holm-corrected $p<.001$). We study how such local sensitivity changes across perturbations, models, and multi-agent interaction, and use TRAILS to connect robustness audits to the scope of scientific claims.
}
\vspace{-1em}
\label{fig:butterfly_effect_illustration}
\end{figure}

% \textbf{LLM social simulations for collective human behavior modeling.}
Until recently, simulating collective human behavior in agent-based models required encoding every social tendency by hand: how an agent decides to cooperate, who it talks to, what it remembers, and how its preferences change. Large language models (LLM) have lifted this constraint. By generating agent behavior directly from natural language, they have opened a new design space for studying a wide range of social phenomena \citep{park2023generativeagents}. The field has advanced rapidly, and these simulations have been studied in a wide range of contexts. Researchers now use LLM-based social simulations to \textit{explore and explain} social mechanisms such as cooperation \citep{akata2025repeatedgames}, opinion dynamics \citep{chuang2024simulating} and social norms emergence \citep{ashery2025emergent}; to serve as \textit{intervention testbeds} for content moderation \citep{liu2025mosaic} and platform design \citep{touzel2025sandboxsocial}; and to \textit{inform real-world decisions} in policy, public health, and epidemiology \citep{chopra2025limits, phfscience2024digitaltwin}. Simulations are moving from exploring what \textit{might} happen to claiming \textit{why} it happens, \textit{when} it will happen, and what should happen next.

% \textbf{A validation gap: Realism without robustness.}
As the stakes have risen, so has the field's attention to validation \citep{anthis2025promising,larooij2025validation, puelmaTouzel2026validationgap}. However, existing efforts focus mostly on whether a simulation \textit{resembles} the world: whether agents behave plausibly \citep{park2023generativeagents}, whether outcomes match empirical patterns \citep{larooij2025validation, yang2024oasis, gu2025echochamber}, whether interaction is grounded in established theory \citep{mou2024hisim, liu2024skepticism}, or whether human evaluators find the results convincing \citep{puelmaTouzel2026validationgap}. Realism and empirical grounding address essential dimensions of validity, but they do not determine whether a simulation is \textit{robust}: its conclusions should remain stable across reasonable alternative configurations, just as empirical findings are expected to survive alternative controls, operationalizations, and sensitivity analyses \citep{simonsohn2020specification, steegen2016increasing}. Without robustness testing, we cannot tell whether a finding reflects the social process being modeled or the design choices used to model it.

% \textbf{Architecture sensitivity and the butterfly effect.}
Robustness matters because LLM-based agents introduce the complexity that traditional rule-based agent-based modeling does not. A simulation can depend on choices about model substrate, persona representation, memory, network initialization, and interaction protocols \citep{zhou2025pimmur}. Choices that appear equivalent to researchers may elicit different decisions from an LLM \citep{salinas2024butterfly, sclar2024quantifying}. A small representational change can alter an agent's decision. In a multi-agent simulation, such decisions can in turn alter subsequent interactions, memories, or social signals through repeated interaction, so the local differences can compound into substantially different macro-level outcomes \citep{bertolotti2020sensitivity}. We use \emph{butterfly effect} \citep{ferrara2024butterfly} to describe this propagation of behavioral sensitivity through interaction. The process can amplify or attenuate initial differences, with effects appearing in both behavioral and collective measures.

To study this phenomenon, we conducted two controlled simulation regimes across four LLM families (Figure~\ref{fig:butterfly_effect_illustration}): a repeated Prisoner's Dilemma comparing fixed-opponent and reciprocal LLM--LLM interaction, and a social-media echo-chamber simulation tracing sensitivity from agent-level choices to collective interaction structure. We varied representation and design choices while holding other components fixed. We found that sensitivity differed across perturbations and models: persona representation substantially shifted cooperation for several models, while other perturbations had smaller or model-dependent effects. In addition, multi-agent interaction further reshapes this sensitivity: reciprocal LLM--LLM interaction amplifies the same persona perturbation for three model families but attenuates it for a fourth, while networked interaction can carry representation sensitivity from interaction targeting to collective network structure. These results show that claims from LLM-based social simulations should match their robustness evidence. We introduce \textbf{TRAILS} (\emph{Taxonomy for Robustness Audits In LLM Simulations}) and the \textbf{Social Simulation Audit Card} for the robustness audits for LLM social simulation.

\section{Related Work}
\label{sec:related_work}

\paragraph{LLM social simulation and the validation gap.}

Agent-based modeling (ABM) provides a bottom-up framework for studying how collective patterns emerge from individual agents and local interactions \citep{schelling1971dynamic,macy2002factors}, while longstanding work highlights challenges in empirical validation, replication, and sensitivity \citep{windrum2007empirical,ormerod2009validation,wilensky2007making,bertolotti2020sensitivity}. LLM-based agents extend this paradigm by generating decisions, communication, and adaptation through natural language \citep{park2023generativeagents}, and have been used to study cooperation and negotiation \citep{akata2025repeatedgames,fontana2025nicer,abdelnabi2023negotiation}, economic and population dynamics \citep{li2024econagent,mi2025mfllm,yang2025twinmarket}, coordinated information operations \citep{orlando2026emergent}, and open-ended social phenomena \citep{gu2025echochamber,wang2025decoding,ren2024socialnorms}. Existing validation efforts examine behavioral plausibility, human evaluation, empirical correspondence, real-world grounding, and theory-guided interaction mechanisms \citep{park2023generativeagents,puelmaTouzel2026validationgap,gu2025echochamber,mou2024hisim,liu2024skepticism}. These approaches assess whether simulations credibly represent their targets, while robustness asks whether their scientific conclusions remain stable across reasonable alternative implementations. Although individual studies include sensitivity checks, systematic analysis across representation, interaction, and design choices remains less developed. Following sensitivity analysis in ABMs \citep{saltelli2008global,windrum2007empirical,bertolotti2020sensitivity} and robustness analysis in empirical research \citep{steegen2016increasing,simonsohn2020specification}, we examine whether conclusions from LLM social simulations persist under controlled changes to simulation design and representation.

\paragraph{From local LLM sensitivity to multi-agent propagation.}

Small changes in LLM inputs can affect individual decisions and potentially propagate through multi-agent interaction. Prior work documents sensitivity to prompt formatting, option ordering, instruction structure, and other surface features \citep{salinas2024butterfly,sclar2024quantifying,pezeshkpour2024sensitivity,razavi2025benchmarking,ju2024sense}, as well as persona construction \citep{li2025llm}, strategic framing, and interaction-history presentation \citep{akata2025repeatedgames,lore2024strategic,fontana2025nicer}. Classical ABM research similarly shows that collective outcomes can depend on initial conditions and modeling assumptions \citep{bertolotti2020sensitivity,windrum2007empirical,wilensky2007making}. In a multi-agent simulation, an LLM output becomes an action, message, memory, or social signal that can enter subsequent interactions. Prior work therefore establishes sensitivity at both the model and system levels, but leaves open how the magnitude of a controlled representation perturbation changes as LLM agents interact. We examine this gap by measuring amplification or attenuation under reciprocal LLM--LLM interaction and by tracing representation sensitivity across behavioral and collective outcomes in networked interaction.

\paragraph{Controlled testbeds for collective sensitivity.}

We study this problem in two established but complementary simulation settings. Repeated Prisoner's Dilemma is a common testbed for strategic behavior and cooperation in LLM agents \citep{akata2025repeatedgames,fontana2025nicer}. Its fixed action space, payoff structure, and interaction horizon provide a controlled setting for comparing the same representation perturbation across fixed-opponent and reciprocal LLM--LLM interaction. Echo chambers provide a complementary open-ended setting in which collective structure emerges from repeated local engagement on a social network. Prior LLM-agent simulations study the roles of network structure, homophily, recommendation and exposure, memory, and interaction dynamics in polarization and echo-chamber formation \citep{gu2025echochamber,wang2025decoding,ferraro2024agent,piao2025emergence}. This setting allows a representation perturbation to be examined first in agents' interaction choices and then in the network formed by those choices. The experiments isolate selected perturbations while holding non-targeted components fixed, allowing sensitivity and its development through interaction to be measured directly. Appendix~\ref{appendix:extended-related-works} provides the broader scope of LLM social simulation and the domain-specific motivation for the perturbations evaluated in each testbed.

\section{Method}
\label{sec:method}

\subsection{Conceptual Framework}
\label{sec:conceptual_framework}

Let $\theta$ denote a social-simulation configuration and $\xi$ a stochastic realization. A controlled perturbation $p$ changes one component of $\theta$ while holding the remaining components fixed, yielding $\theta^{(p)}$. For an outcome $Y$, we define the signed perturbation effect and its sensitivity magnitude as
\begin{equation}
\Delta_Y(p)
=
\mathbb{E}_{\xi}
\left[
Y(\theta^{(p)},\xi)-Y(\theta,\xi)
\right],
\qquad
S_Y(p)=|\Delta_Y(p)|.
\label{eq:sensitivity}
\end{equation}
Sensitivity measures the change associated with a controlled perturbation.

A perturbation can change agent behavior, which changes the interaction context available to other agents. We use \emph{butterfly effect} to describe the resulting interaction-mediated propagation of sensitivity:
\begin{equation}
\boxed{
p
\;\longrightarrow\;
\underbrace{\text{behavioral sensitivity}}_{\text{agent level}}
\;\xrightarrow{\text{interaction}}\;
\underbrace{\text{collective divergence}}_{\text{system level}}
}
\label{eq:butterfly_process}
\end{equation}
When local and collective sensitivity are measured on a common outcome scale, interaction amplification corresponds to $S_{\mathrm{collective}}(p)>S_{\mathrm{local}}(p)$. The corresponding difference in effect magnitude is positive under amplification and negative under attenuation. When propagation spans outcomes measured at different levels, we characterize the stages separately. We distinguish \emph{representation perturbations}, which vary how a fixed simulation specification is presented to the LLM, from \emph{design perturbations}, which vary components of the simulated system itself; Section~\ref{sec:trails} generalizes this distinction in TRAILS.

% \subsection{Two Complementary Simulation Regimes}
% \label{sec:simulation_regimes}

% We study two complementary interaction regimes. Repeated Prisoner's Dilemma (PD) provides a controlled setting for cooperation under repeated dyadic interaction \citep{akata2025repeatedgames,fontana2025nicer}, allowing perturbation sensitivity to be compared with and without reciprocal LLM interaction. Social-media echo-chamber (EC) simulation captures networked interaction in which agents repeatedly choose whom and what to engage with \citep{gu2025echochamber,wang2025decoding,piao2025emergence}, allowing sensitivity to be traced from agent-level choices to realized interaction structure. Across both regimes, we vary selected \emph{representation perturbations}, which change how simulation information is presented to the LLM, and \emph{design perturbations}, which change components of the simulated interaction process, while holding non-targeted components fixed. Full perturbation specifications are provided in Appendices~\ref{appendix:pd-perturbations} and~\ref{appendix:ec-perturbations}; Section~\ref{sec:trails} generalizes these dimensions into the TRAILS audit taxonomy.

\subsection{Repeated Prisoner's Dilemma}
\label{sec:pd_method}

\noindent \textbf{Setup.}
We use a 10-round repeated Prisoner's Dilemma with actions Cooperate ($C$) and Defect ($D$) and payoffs $(C,C)=(3,3)$, $(C,D)=(0,5)$, $(D,C)=(5,0)$, and $(D,D)=(1,1)$. In the single-agent regime, one LLM plays against four fixed policies, $\mathcal{O}=\{\text{Tit-for-Tat},\text{Random},\text{Always-Cooperate},\text{Always-Defect}\}$; in the two-agent regime, two matched LLM agents play each other. Each condition contains $N=30$ runs.

\noindent \textbf{Perturbations and measures.}
We vary three representation dimensions: persona format (\textsc{Plain}, \textsc{Descriptive}, \textsc{Tabular}), game framing (\textsc{Canonical}, \textsc{Moralized}, \textsc{Risk}), and memory representation (table or narrative history, with or without summary statistics). We measure cooperation rate as the primary outcome and average per-round payoff as a secondary outcome. Full configurations, prompt constructions, and metric definitions are provided in Appendix~\ref{appendix:pd-details}.

\noindent \textbf{Interaction amplification.}
For model $m$ and representation contrast $p=(c_0,c_1)$, let
\[
\Delta_{m,p,o}
=
\mathbb{E}(Y\mid c_1,m,o)
-
\mathbb{E}(Y\mid c_0,m,o)
\]
denote the signed perturbation effect against fixed opponent $o\in\mathcal{O}$. We summarize sensitivity across the four benchmark contexts as
\[
S^{\mathrm{fixed}}_{m,p}
=
\frac{1}{|\mathcal{O}|}
\sum_{o\in\mathcal{O}}
|\Delta_{m,p,o}|.
\]
Under LLM--LLM interaction,
\[
S^{\mathrm{LLM}}_{m,p}
=
\left|
\mathbb{E}(Y\mid c_1,m,\mathrm{LLM})
-
\mathbb{E}(Y\mid c_0,m,\mathrm{LLM})
\right|.
\]
The interaction effect gap is
\[
A_{m,p}
=
S^{\mathrm{LLM}}_{m,p}
-
S^{\mathrm{fixed}}_{m,p}.
\]
We refer to $A_{m,p}>0$ as \emph{amplification} and $A_{m,p}<0$ as \emph{attenuation}; $A_{m,p}=0$ indicates equal sensitivity across interaction regimes.

\subsection{Social-Media Echo Chamber}
\label{sec:ec_method}

\noindent \textbf{Setup.}
We simulate $N=100$ agents for $T=15$ rounds on a fixed social network. Each agent has a persona and fixed stance $s_i\in\{-2,-1,0,1,2\}$. Activated agents observe recent neighbor posts and choose among \textsc{Post}, \textsc{Repost}, \textsc{Reply}, and \textsc{Silent}; reposts and replies additionally select a visible target. Stances remain fixed, so collective outcomes capture the interaction structure that emerges conditional on the initial stance distribution.

\noindent \textbf{Perturbations and measures.}
We vary persona representation (\textsc{Bullet} vs.\ \textsc{Prose}) and five design dimensions: input-network homophily, hub assignment, activation probability, memory window, and recommendation feed size. We measure agent-level interaction targeting and collective network structure using stance assortativity and weighted same-group edge ratio. Full configurations, implementation details, and metric definitions are provided in Appendix~\ref{appendix:ec-details}.

\noindent \textbf{Sensitivity across interaction levels.}
For perturbation $p=(c_0,c_1)$, let $q_t^{(c)}$ denote the cross-group targeting rate and $Y_t^{(c)}$ a collective network outcome under condition $c$, with $Y\in\{r,H\}$. We trace sensitivity across interaction levels using
\begin{equation}
\begin{aligned}
\Delta_{\mathrm{target},t}(p)
&=
q_t^{(c_1)}-q_t^{(c_0)},\\
\Delta_{Y,t}(p)
&=
Y_t^{(c_1)}-Y_t^{(c_0)},
\qquad Y\in\{r,H\},
\end{aligned}
\label{eq:ec_sensitivity}
\end{equation}
with corresponding sensitivity magnitudes
$S_{\mathrm{target},t}(p)=|\Delta_{\mathrm{target},t}(p)|$
and
$S_{Y,t}(p)=|\Delta_{Y,t}(p)|$.

We additionally compare action-type and target agreement across paired runs to locate behavioral sensitivity in action selection or interaction targeting. We trace propagation from agent-level behavioral sensitivity to collective network divergence. Formal metric definitions are provided in Appendix~\ref{appendix:ec-details}.

\paragraph{Models and statistical analysis.}

We execute both experiments on four model families: \texttt{gpt-5.2}, \texttt{claude-haiku-4-5}, 
\texttt{gemini-2.5-flash}, and \texttt{deepseek}\footnote{For the DeepSeek family, 
we use \texttt{deepseek-v3.2} in the Prisoner's Dilemma (PD) experiments and 
\texttt{deepseek-v4-flash} in the echo-chamber (EC) experiments.}. The simulation run is the unit of analysis, with $N=30$ independent runs per condition unless otherwise specified. For paired comparisons, conditions share the same simulation seed. We report effect estimates with 95\% confidence intervals and use two-sided tests with Holm correction for multiple pairwise comparisons. Full prompts and experimental configurations are provided in Appendices~\ref{appendix:simulation-details} and~\ref{appendix:prompts}.

\begin{figure}[t]
    \centering
    \includegraphics[width=\textwidth]{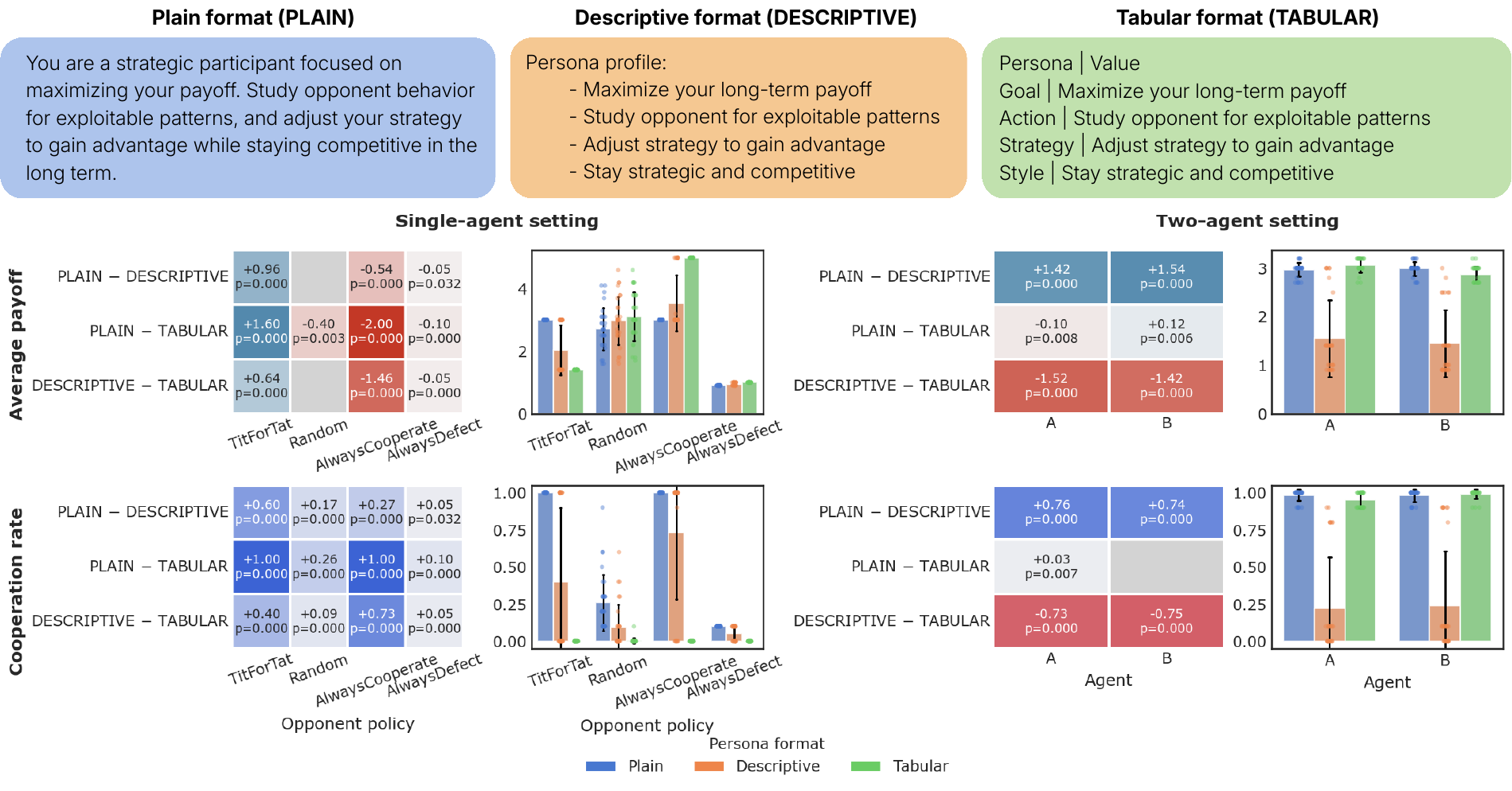}
    % \captionsetup{font=footnotesize}
    \caption{\textbf{Effect of persona format on Prisoner's Dilemma outcomes} 
(\texttt{gpt-5.2}). Left: single-agent (fixed-opponent); right: two-agent (LLM--LLM). Bars show means with SD error bars; heatmaps show pairwise differences, with color indicating significance ($p<.05$) and gray non-significance.}
    \label{fig:persona_format_panel}
\end{figure}

\section{Results}
\label{sec:results}

\subsection{Sensitivity Varies Across Perturbations and Models}

\definecolor{sigone}{RGB}{252,225,230}   % p < .05
\definecolor{sigtwo}{RGB}{247,181,194}   % p < .01
\definecolor{sigthree}{RGB}{238,128,151} % p < .001

\noindent \textbf{Representation choices can substantially shift simulated outcomes.}
Figure~\ref{fig:persona_format_panel} shows a representative case in the repeated Prisoner's Dilemma. For \texttt{gpt-5.2}, changing only the representation of the matched strategic persona produces a large shift in the two-agent (LLM--LLM) setting: cooperation falls from $98.2\%$ under \textsc{Plain} to $23.0\%$ under \textsc{Descriptive}, a decrease of $75.2$ percentage points (95\% CI $[62.0,88.3]$). Average payoff decreases by $1.48$ points per round over the same contrast ($p<.001$ for both outcomes). The same representation contrast produces different effects in the single-agent (fixed-opponent) setting, motivating the interaction-regime analysis in Section~\ref{sec:results_amplification}. Complete directional contrasts are reported in Appendix~\ref{appendix:extended-results}.

\noindent \textbf{Representation sensitivity varies across perturbations and models.}
Persona format produces significant two-agent cooperation shifts across all four model families, with the largest within-dimension contrast ranging from $7.7$ percentage points for \texttt{deepseek-v3.2} to $77.7$ percentage points for \texttt{claude-haiku-4-5} (Table~\ref{tab:sensitivity-summary}). Game framing produces significant cooperation shifts for a subset of models, while memory representation yields a significant maximum cooperation contrast only for \texttt{gpt-5.2}. The observed sensitivity therefore depends on both the representation dimension and the model family.

\begin{table*}[t]
\centering
\caption{\textbf{Outcome sensitivity across perturbations and models.} 
$\Delta$ denotes the difference in outcome means between two perturbation conditions.}
\label{tab:sensitivity-summary}
\vspace{0.5em}

\scriptsize
\setlength{\tabcolsep}{4.5pt}
\renewcommand{\arraystretch}{1.12}

\begin{tabular}{llcccc}
\toprule
\textbf{Perturbation}
& \textbf{Metric}
& \textit{GPT-5.2}
& \textit{Claude Haiku 4.5}
& \textit{Gemini 2.5 Flash}
& \textit{DeepSeek} \\
\midrule

\multicolumn{6}{c}{\textbf{Repeated Prisoner's Dilemma (Two-Agent): $\max|\Delta|$}} \\
\midrule

\multirow{2}{*}{Persona format$^{\mathrm{R}}$}
& Average payoff
& \cellcolor{sigthree}1.48***
& \cellcolor{sigthree}1.49***
& \cellcolor{sigthree}0.79***
& \cellcolor{sigtwo}0.14** \\
& Cooperation rate (pp)
& \cellcolor{sigthree}75.2***
& \cellcolor{sigthree}77.7***
& \cellcolor{sigthree}36.0***
& \cellcolor{sigtwo}7.7** \\

\addlinespace[2pt]

\multirow{2}{*}{Game framing$^{\mathrm{R}}$}
& Average payoff
& \cellcolor{sigthree}0.42***
& \cellcolor{sigone}0.09*
& 0.19
& \cellcolor{sigthree}0.15*** \\
& Cooperation rate (pp)
& \cellcolor{sigthree}19.3***
& \cellcolor{sigone}3.3*
& 9.0
& \cellcolor{sigthree}8.8*** \\

\addlinespace[2pt]

\multirow{2}{*}{Memory representation$^{\mathrm{R}}$}
& Average payoff
& \cellcolor{sigthree}0.10***
& 0.05
& 0.29
& 0.01 \\
& Cooperation rate (pp)
& \cellcolor{sigthree}6.2***
& 2.7
& 13.8
& 1.0 \\

\midrule

\multicolumn{6}{c}{\textbf{Social-Media Echo Chamber: $|\Delta|$}} \\
\midrule

\multirow{2}{*}{Persona representation$^{\mathrm{R}}$}
& Stance assortativity
& \cellcolor{sigthree}0.194***
& 0.002
& \cellcolor{sigthree}0.035***
& 0.023 \\
& Same-group edge ratio
& \cellcolor{sigthree}0.051***
& \cellcolor{sigone}0.010*
& \cellcolor{sigthree}0.023***
& \cellcolor{sigthree}0.072*** \\

\addlinespace[2pt]

\multirow{2}{*}{Activation probability$^{\mathrm{D}}$}
& Stance assortativity
& \cellcolor{sigthree}0.068***
& \cellcolor{sigthree}0.103***
& \cellcolor{sigthree}0.108***
& \cellcolor{sigthree}0.107*** \\
& Same-group edge ratio
& \cellcolor{sigthree}0.032***
& \cellcolor{sigthree}0.060***
& \cellcolor{sigthree}0.061***
& \cellcolor{sigthree}0.070*** \\

\addlinespace[2pt]

\multirow{2}{*}{Memory window$^{\mathrm{D}}$}
& Stance assortativity
& 0.009
& \cellcolor{sigthree}0.063***
& \cellcolor{sigthree}0.080***
& \cellcolor{sigthree}0.092*** \\
& Same-group edge ratio
& 0.011
& \cellcolor{sigthree}0.039***
& \cellcolor{sigthree}0.038***
& \cellcolor{sigthree}0.058*** \\

\addlinespace[2pt]

\multirow{2}{*}{Recommendation feed size$^{\mathrm{D}}$}
& Stance assortativity
& 0.001
& 0.010
& \cellcolor{sigone}0.015*
& 0.008 \\
& Same-group edge ratio
& 0.006
& \cellcolor{sigtwo}0.012**
& \cellcolor{sigtwo}0.009**
& 0.006 \\

\bottomrule
\end{tabular}

\vspace{2pt}

\begin{minipage}{0.96\textwidth}
\tiny
\textit{Note:}
PD reports the maximum absolute effect across prespecified pairwise contrasts within each representation dimension; EC reports the absolute effect of the binary contrast.
PD cooperation effects are in percentage points; other outcomes use their native units.
Superscripts denote perturbation type: $\mathrm{R}$ = representation-level (TRAILS-R) and $\mathrm{D}$ = design-level (TRAILS-D).
Shading denotes significance:
*** $p<.001$, ** $p<.01$, * $p<.05$; unshaded cells are not significant.
For DeepSeek, PD experiments use \texttt{deepseek-v3.2}, whereas EC experiments use \texttt{deepseek-v4-flash}.
\end{minipage}
\vspace{-1em}
\end{table*}

\noindent \textbf{Sensitivity is similarly selective in networked interaction.}
In the echo-chamber simulation, persona representation significantly changes stance assortativity for two of the four models, with the largest effect ($0.194$) observed for \texttt{gpt-5.2}, and significantly changes the same-group edge ratio across all four models (Table~\ref{tab:sensitivity-summary}). Among design perturbations, activation probability significantly affects both outcomes across all four models, while memory window does so for three. Recommendation feed size produces smaller, more selective effects. Sensitivity in the networked simulation therefore varies across representation and design choices, models, and collective outcomes. Signed design-perturbation contrasts and confidence intervals are reported in Appendix~\ref{appendix:extended-results}, Table~\ref{tab:ec-design-full}.
% Complete directional effects are reported in Appendix~\ref{appendix:extended-results}.

Across both regimes, configuration-level sensitivity varies across perturbations and models. We next examine how multi-agent interaction reshapes representation sensitivity, through amplification or attenuation in the Prisoner's Dilemma and across behavioral and collective levels in the echo-chamber simulation.

\subsection{Multi-Agent Interaction Reshapes Representation Sensitivity}
\label{sec:results_amplification}

The preceding results establish configuration-level sensitivity across representation and design perturbations. We focus the interaction analysis on persona representation, which produces clear sensitivity in both simulation regimes and provides a common perturbation for examining how local differences develop through interaction. The repeated Prisoner's Dilemma permits direct comparison under fixed-opponent and LLM--LLM interaction, while the echo-chamber simulation traces sensitivity across interaction targeting and collective network structure.

\begin{figure}[t]
    \centering
    \includegraphics[width=\textwidth]{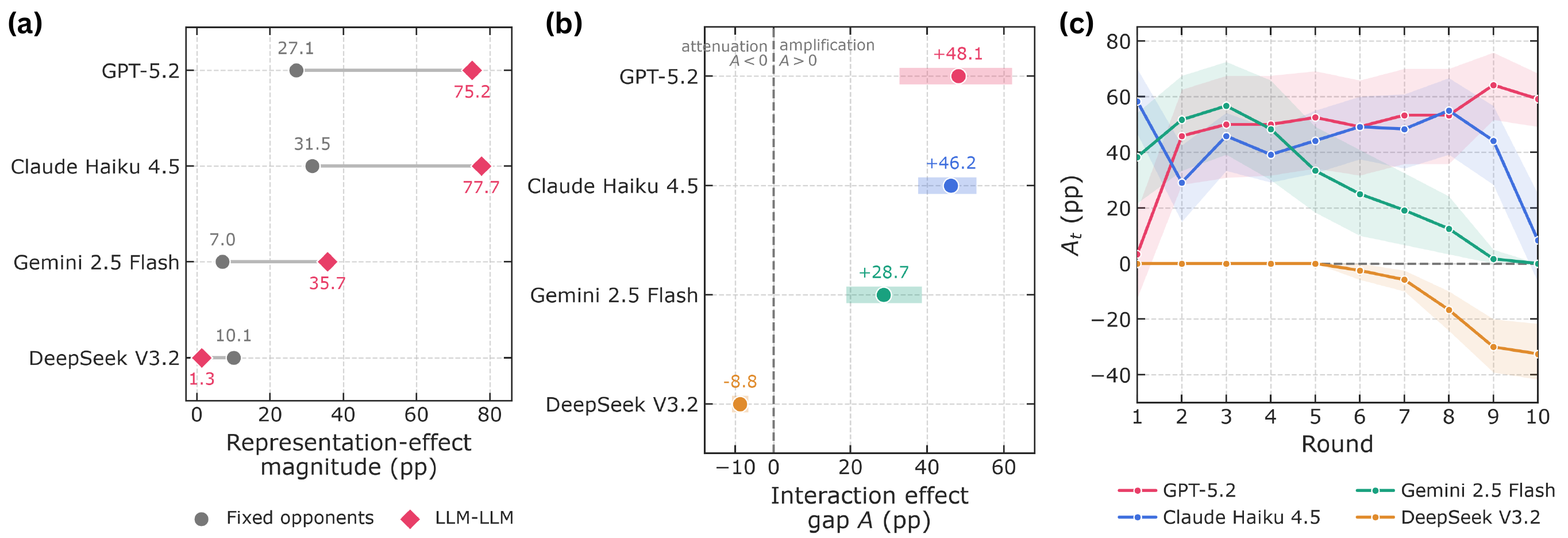}
    % \captionsetup{font=footnotesize}
    \caption{\textbf{Interaction reshapes persona-representation sensitivity in the repeated Prisoner's Dilemma.}
\textsc{Plain} vs.\ \textsc{Descriptive}, $N=30$ paired seeds.
(a) Cooperation-rate sensitivity under fixed-opponent and LLM--LLM interaction.
(b) Interaction effect gap $A=S^{\mathrm{LLM}}-S^{\mathrm{fixed}}$; positive values indicate amplification and negative values attenuation.
(c) Round-wise effect gap $A_t$.
Intervals and bands show paired-bootstrap 95\% CIs.}
    \label{fig:PD_amplification}
    \vspace{-1em}
\end{figure}

\noindent \textbf{Reciprocal interaction can amplify or attenuate representation sensitivity.}
For the \textsc{Plain}$\rightarrow$\textsc{Descriptive} persona contrast, \texttt{gpt-5.2} sensitivity increases from $S^{\mathrm{fixed}}=0.271$ against fixed opponents to $S^{\mathrm{LLM}}=0.752$ under LLM--LLM interaction, yielding $A=0.481$ (95\% CI $[0.327,0.622]$; Figure~\ref{fig:PD_amplification}). The interaction effect is also positive for \texttt{claude-haiku-4-5} ($A=0.462$) and \texttt{gemini-2.5-flash} ($A=0.287$), while \texttt{deepseek-v3.2} shows attenuation ($A=-0.088$). Round-wise estimates show different temporal trajectories across models. Reciprocal interaction therefore changes the magnitude and development of representation sensitivity, with the direction of modulation depending on the model. Complete estimates and confidence intervals are reported in Appendix~\ref{appendix:extended-results}.

\begin{figure}[t]
    \centering
    \includegraphics[width=\textwidth]{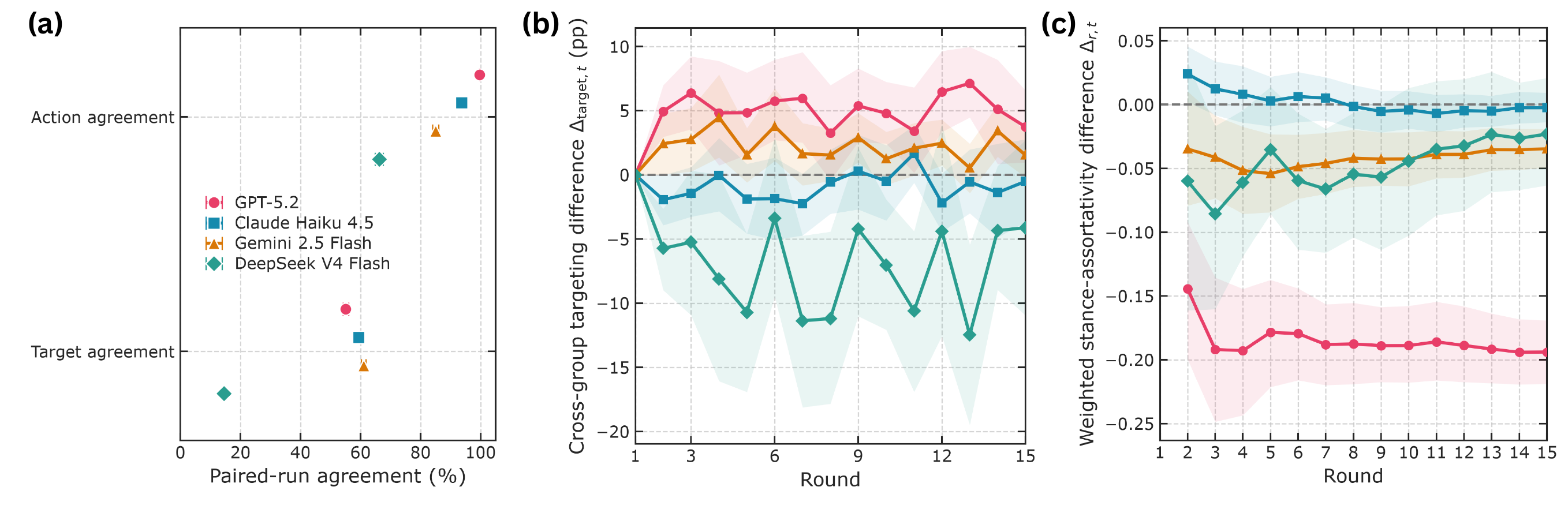}
    % \captionsetup{font=footnotesize}
    \caption{\textbf{Persona-representation sensitivity across interaction levels in the echo-chamber simulation.}
\textsc{Prose} vs.\ \textsc{Bullet}, $N=30$ paired seeds.
(a) Agreement in action type and interaction target.
(b) Cross-group targeting difference $\Delta_{\mathrm{target},t}$; positive values indicate more cross-group targeting under \textsc{Prose}.
(c) Weighted stance-assortativity difference $\Delta_{r,t}$; negative values indicate lower assortativity under \textsc{Prose}.
Error bars and bands show 95\% CIs.}
    \label{fig:EC_amplification}
    \vspace{-1em}
\end{figure}

\noindent \textbf{Representation sensitivity can extend from targeting behavior to collective structure.}
In the echo-chamber simulation, persona representation changes interaction targets even when action selection remains comparatively stable (Figure~\ref{fig:EC_amplification}). For \texttt{gpt-5.2}, action agreement is $99.7\%$, while target agreement is $55.1\%$. \textsc{Prose} increases cross-group targeting by $5.15$ percentage points (95\% CI $[4.22,6.07]$) and is associated with a $0.194$ lower cumulative stance assortativity by round $15$. The pattern varies across models: \texttt{gemini-2.5-flash} shows increased cross-group targeting ($+2.31$ pp) and lower assortativity ($-0.035$), while the final assortativity differences for \texttt{claude-haiku-4-5} and \texttt{deepseek-v4-flash} are not statistically significant. Representation sensitivity can therefore appear at both behavioral and collective levels, with the relationship between the two varying across models. Complementary design-perturbation trajectories show differences in both interaction targeting and cumulative interaction structure, with patterns varying across perturbations and models. Complete estimates and confidence intervals are reported in Appendix~\ref{appendix:extended-results}.

The two regimes capture complementary forms of interaction-mediated sensitivity. In repeated dyadic interaction, a perturbation effect on a common outcome can be amplified or attenuated. In networked interaction, representation sensitivity can appear across behavioral and collective levels.

% \subsection{Sensitivity to Context-Relevant Design Choices}

\section{From Architecture Sensitivity to Robustness Audits}
\label{sec:trails}

\paragraph{TRAILS.}
LLM social simulations can be sensitive to both design-level perturbations, which change the model, network, or interaction protocol, and representation-level perturbations, which change how a fixed setup is presented through instructions, personas, labels, or memory formats. Our experiments found sensitivity to both, with effects varying across perturbations and models. We introduce \textbf{TRAILS} (\textit{Taxonomy for Robustness Audits In LLM Simulations}) to organize these choices into representation (TRAILS-R) and design (TRAILS-D) audit dimensions (Table~\ref{tab:trails}), given the levels identified in Appendix~\ref{appendix:extended-related-works}. \textbf{TRAILS-D} organizes changes to the substantive simulation setup into eight dimensions across micro, meso, and macro levels. \textbf{TRAILS-R} organizes changes to how a matched simulation specification or context is presented to LLM agents into five dimensions. These choices can affect how agents interpret their inputs. Appendix~\ref{appendix:taxonomy} provides the examples of TRAILS-R and TRAILS-D. TRAILS covers the simulation system; measurement and evaluation sensitivity falls within the downstream evaluation pipeline. 

\begin{table*}[t]
\caption{
\textbf{TRAILS: Taxonomy for Robustness Audits in LLM Simulations.}
TRAILS distinguishes representation-level perturbations (TRAILS-R), which vary how a fixed simulation specification or context is presented to LLM agents, from design-level perturbations (TRAILS-D), which vary components of the simulated system itself.
\textbf{Bolded examples} indicate perturbations examined in our case studies.
}
\label{tab:trails}

\centering
\scriptsize
\setlength{\tabcolsep}{4pt}
\renewcommand{\arraystretch}{1.08}
\vspace{0.8em}

\begin{tabular}{
p{0.06\textwidth}
p{0.20\textwidth}
p{0.45\textwidth}
p{0.20\textwidth}
}
\toprule
\textbf{Level} &
\textbf{Dimension} &
\textbf{Description} &
\textbf{Examples} \\
\midrule

\multicolumn{4}{l}{\textbf{TRAILS-R: Representation-level perturbations}} \\
\midrule

& Representational format
& Alternative encodings of a matched agent or task specification, including prose, lists, tables, and structured formats; \textbf{persona format}
& \citet{salinas2024butterfly,sclar2024quantifying}. \\

& Instruction hierarchy
& Instruction order, placement, grouping, and authority across prompt components
& \citet{salinas2024butterfly,sclar2024quantifying}. \\

& Linguistic framing
& Alternative wording or labels while other simulation components are held fixed; \textbf{game-instruction framing}
& \citet{salinas2024butterfly,sclar2024quantifying}. \\

& Context representation
& Alternative representations or encodings of the same interaction history and contextual information; \textbf{memory representation}
& \citet{park2023generativeagents,salinas2024butterfly}. \\

& Interaction sequencing
& Presentation or ordering of otherwise matched interactions
& \citet{mou2024hisim}. \\

\midrule

\multicolumn{4}{l}{\textbf{TRAILS-D: Design-level perturbations}} \\
\midrule

\multirow{4}{*}{\textbf{Micro}}
& Model substrate
& Model family, model variant, inference settings, and decoding parameters
& \citet{zhou2024sotopia}. \\

& Agent specification
& Persona attributes, goals, beliefs, demographic or behavioral profiles, and population grounding
& \citet{park2023generativeagents,li2025llm}. \\

& Internal state \& cognition
& Belief and preference updating, reasoning procedures, reflection rules, and internal-state representations
& \citet{liu2024skepticism,wang2023humanoid}. \\

& Memory \& temporality
& Memory length, retrieval, summarization, forgetting, and temporal granularity
& \citet{park2023generativeagents,wang2025userbehavior}. \\

\midrule

\multirow{2}{*}{\textbf{Meso}}
& Interaction protocol
& Turn structure, available actions, visibility, activation, and rules governing who can interact with whom; \textbf{activation probability}
& \citet{mou2024hisim,yang2024oasis}. \\

& Intervention design
& Intervention type, target, timing, frequency, delivery mechanism, and framing
& \citet{liu2025mosaic,liu2024skepticism}. \\

\midrule

\multirow{2}{*}{\textbf{Macro}}
& Environment structure
& Network topology, exposure and recommendation rules, institutional structure, and platform affordances; \textbf{network homophily, hub assignment, recommendation feed size}
& \citet{yang2024oasis,gu2025echochamber,piao2025agentsociety}. \\

& Population \& scale
& Population size, composition, heterogeneity, agent assignment, and real- versus synthetic-population construction
& \citet{yang2024oasis,wang2025yulanonesim,piao2025agentsociety}. \\

\bottomrule
\end{tabular}
\end{table*}

\paragraph{The Social Simulation Audit Card.}
We introduce the Social Simulation Audit Card to operationalize TRAILS for a stated scientific claim. The card guides researchers through five steps: stating the claim, mapping perturbations to TRAILS and context relevance, registering perturbations and preservation criteria, reporting sensitivity, and defining the claim's scope. Context relevance is claim-specific: context-relevant perturbations represent substantively meaningful conditions of the modeled phenomenon, whereas context-irrelevant perturbations probe implementation dependence. For example, network homophily and hub assignment are context-relevant structural conditions in the echo-chamber simulation; their sensitivity results are reported in Appendix~\ref{appendix:extended-results}. Accordingly, context-relevant sensitivity helps scope where a claim holds, while substantial sensitivity to context-irrelevant choices indicates implementation dependence. Outcomes are classified as \textit{Robust}, \textit{Scope-limited}, \textit{Implementation-sensitive}, or \textit{Inconclusive}, bounded by the audited conditions. Appendix~\ref{appendix:audit-card} provides the full procedure and case-study audits.

\begin{figure}[t]
    \centering
    \includegraphics[width=\textwidth]{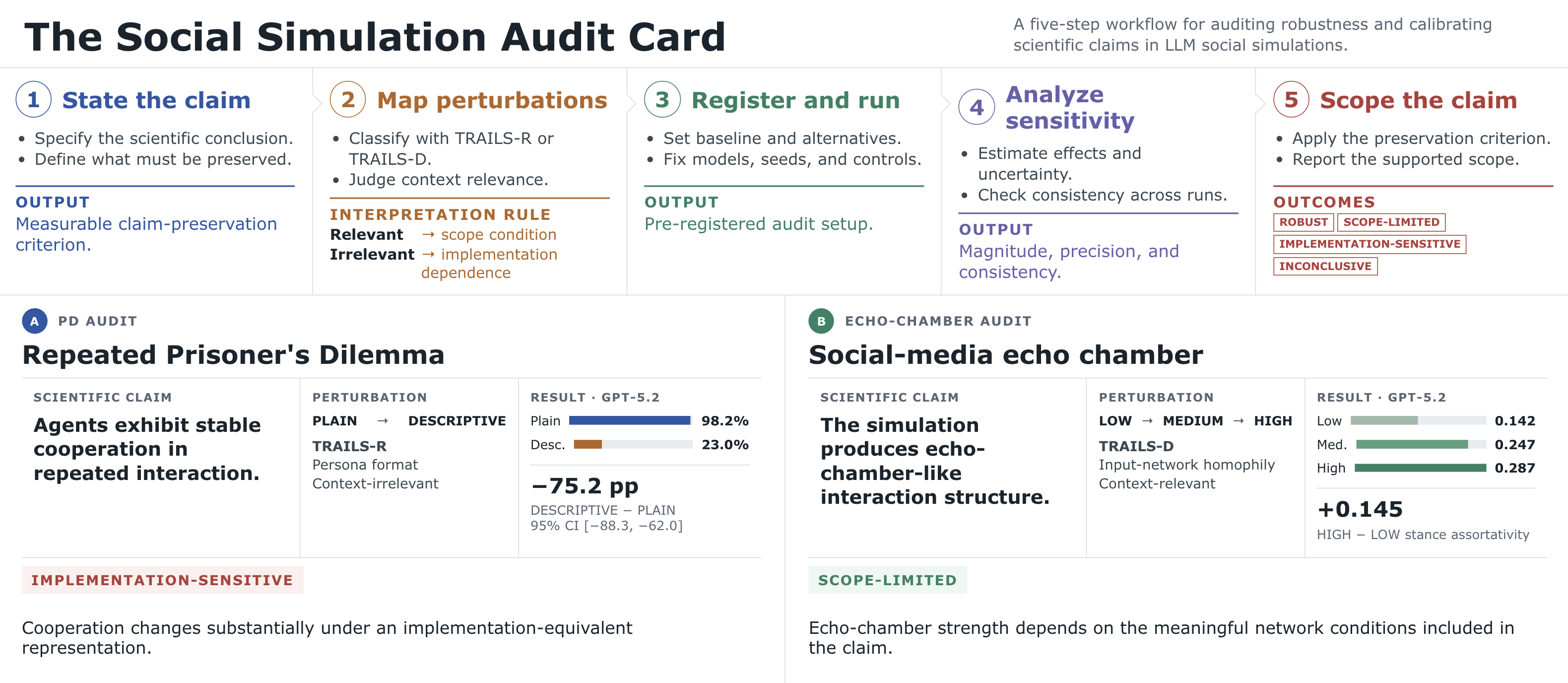}
    % \captionsetup{font=footnotesize}
    \caption{\textbf{The Social Simulation Audit Card.}
The five-step workflow connects perturbation selection and sensitivity
analysis to the scope of scientific claims. GPT-5.2 examples illustrate
two interpretations of sensitivity: a context-irrelevant persona-format
change reveals implementation dependence in the repeated Prisoner's
Dilemma, whereas context-relevant network homophily bounds the conditions
under which echo-chamber-like interaction structure is supported.}
    \label{fig:audit_card}
\end{figure}

\section{Conclusions}

LLM social simulations can be sensitive to choices that are often treated as interchangeable or left unaudited. Across repeated Prisoner's Dilemma and echo-chamber simulations, we find that sensitivity varies across perturbations, models, and outcomes. Persona representation produces substantial shifts in several settings, while other representation and design perturbations have smaller or more selective effects. Multi-agent interaction further reshapes this sensitivity. In repeated dyadic interaction, representation effects can be amplified or attenuated under reciprocal LLM--LLM interaction; in networked interaction, representation sensitivity appears in interaction targeting and, for some models, in collective network structure. These results indicate that robustness observed under one configuration or model does not establish robustness of the resulting scientific claim.

Our findings are limited to two simulation regimes, selected perturbations, four model families, and fixed interaction horizons, so the observed sensitivity patterns should not be assumed to generalize to other settings. The echo-chamber experiment holds agent stances fixed and therefore characterizes interaction structure rather than endogenous opinion change. Its behavioral and collective analyses identify cross-level patterns of sensitivity but do not establish a causal pathway from targeting differences to network divergence. More broadly, robustness to implementation choices addresses only one dimension of simulation validity; correspondence between simulated agents and human behavior requires separate empirical validation.

The broader implication concerns the scope of scientific inference from LLM social simulations. Robustness audits should identify which conclusions persist across plausible implementations, which depend on substantively meaningful conditions, and which remain sensitive to implementation choices. TRAILS and the Social Simulation Audit Card provide a vocabulary and reporting structure for conducting such audits, with the required coverage depending on the claim and simulation context. Scientific claims from simulated societies can then be scoped to the configurations, models, and interaction conditions for which robustness evidence is available.

\bibliography{iclr2027_conference}
\bibliographystyle{iclr2027_conference}

\section*{Relationship to Earlier Work}
\addcontentsline{toc}{section}{Relationship to Earlier Work}

An earlier position paper introduced the robustness-audit perspective and TRAILS framework, with selected results from the same simulation settings serving as motivating examples \citep{ye2026stop}. The present study extends the empirical analysis through systematic sensitivity comparisons across perturbations and models, explicit measurement of interaction amplification and attenuation, and analysis of behavioral and collective outcomes in networked interaction.

\newpage
\appendix

% \section{Appendix}
\section{TRAILS (Taxonomy for Robustness Audits In LLM Simulations)}
\label{appendix:taxonomy}

\begin{table}[htbp]
\caption{TRAILS-D: Design-level perturbations}
\label{tab:trails_d}
\centering
\footnotesize
\begin{tabular}{p{0.1\linewidth} p{0.24\linewidth} p{0.52\linewidth}}
\toprule
\textbf{Level} & \textbf{Dimension} & \textbf{What to perturb} \\
\midrule

\textbf{Micro} 
& Model substrate 
& Model family, model size, base vs. instruction-tuned models, alignment, temperature, top-$p$, random seed, safeguards. \\

& Agent specification 
& Demographics, ideology, personality, goals, prior beliefs, real-data-grounded vs. synthetic personas, relationship attributes. \\

& Internal state and cognition
& Beliefs, attitudes, emotions, needs, moral values, reasoning style, reflection rules, belief updating, preference updating. \\

& Memory and temporality
& last-$k$ memory, summarized memory, episodic memory, retrieval rules, forgetting, reflection frequency, temporal granularity. \\

\midrule

\textbf{Meso} 
& Interaction protocol 
& Turn order, fixed or free turns, dyadic or group interaction, available actions \\

& Intervention design
& Moderator type, target selection, timing, frequency, rule-based vs. LLM-based intervention. \\

\midrule

\textbf{Macro} 
& Environment structure 
& Network topology, feed ranking, recommendation system, moderation rules, platform affordances, institutional rules. \\

& Population and scale 
& Number of agents, demographic distribution, ideological balance, heterogeneity, real vs. synthetic population construction. \\

\bottomrule
\end{tabular}
\end{table}

\begin{table}[htbp]
\caption{TRAILS-R: Representation-level perturbations}
\label{tab:trails_r}
\centering
\footnotesize
\begin{tabular}{p{0.24\linewidth} p{0.24\linewidth} p{0.45\linewidth}}
\toprule
\textbf{Category} & \textbf{Perturbation} & \textbf{Example} \\
\midrule

\textbf{Representational format} 
& Formatting 
& Narrative prose vs. bullet points; paragraph vs. table; bolded keywords; numbered lists. \\

& Delimiter choice 
& Conversation history enclosed in code blocks, XML tags, JSON, or plain text. \\

& Output schema 
& Free-form paragraph vs. short reply vs. JSON with rationale. \\

\midrule

\textbf{Instruction hierarchy} 
& Instruction order 
& “Update your opinion, then write a reply” vs. “Write a reply, then update your opinion.” \\

& Role/message placement 
& The same instruction placed in the system message vs. the user message. \\

\midrule

\textbf{Linguistic framing} 
& Agent naming 
& Agent A/B vs. Alice/Bob. \\

& Persona wording 
& ``Left'' vs. ``Democrat''; ``Right'' vs. ``Republican.'' \\

& Moderation framing 
& Same intervention framed as a warning, suggestion, community reminder, or bridge statement. \\

& Few-shot examples 
& Same examples framed differently. \\

\midrule

\textbf{Context representation} 
& Memory representation 
& Same history shown as transcript, JSON list, bullet summary. \\

& Context window 
& Full history vs. last three turns; same content with early background removed. \\

& Message metadata 
& Adding timestamps, usernames. \\

\midrule

\textbf{Interaction sequencing} 
& Initial seed turn 
& Same stance, but a slightly different opening statement. \\

& Turn order 
& Which agent speaks first. \\

% & Stopping rule 
% & Fixed number of turns vs. stopping after reaching a threshold. \\

% \midrule

% \textbf{Generation interface} 
% & Output length and stopping 
% & Maximum tokens; stop sequence; required brevity; one-sentence reply vs. unconstrained response. \\

\bottomrule
\end{tabular}
\end{table}

\newpage
\section{Extended Related Works}
\label{appendix:extended-related-works}

% In this section, we will discuss the opportunities and validation gap in using LLMs in social simulations for collective human behavior.

% \subsection{Simulation can prove or disprove scientific claim without robust audits}

% assumption vs control conditions

% % lack of standards + instable simulation can prove or disprove any claims.

\subsection{LLM-based social simulations}
\paragraph{Scope.} Following work on generative social simulation and its validation \citep{puelmaTouzel2026validationgap,anthis2025promising,larooij2025validation}, we focus on collective human behavior arising from interacting LLM agents. This scope connects individual decisions, communication, and memory to collective outcomes through a shared social environment. We distinguish it from individual or user simulations without inter-agent interaction \citep{ren2024bases,yu2024large}, simulations centered on non-human or aggregate actors, and multi-agent systems designed primarily for task completion \citep{hong2023metagpt,qian2023chatdev}.

\paragraph{Social simulation scenarios.}
Our two testbeds capture complementary interaction settings. In \textbf{goal-structured scenarios}, explicit incentives and rules organize strategic interaction, as in cooperation and negotiation \citep{akata2025repeatedgames,abdelnabi2023negotiation}. In \textbf{open-ended scenarios}, repeated communication and engagement produce collective patterns, including echo chambers \citep{gu2025echochamber,wang2025decoding}, social norms \citep{ren2024socialnorms}, and coordinated information operations \citep{orlando2026emergent}. This distinction motivates comparing a fixed strategic game with networked social interaction rather than treating either as representative of all social simulations.

\paragraph{Factors shaping audit strength.} Audit requirements depend on the claim, the simulation's openness, and the stakes of its use. Exploratory findings require checks of prompt, seed, and model dependence; mechanism claims require plausible alternative specifications; and intervention or policy claims require broader coverage and greater precision. Open-ended simulations expose more choices about communication, memory, and platform feedback, expanding the perturbation space. Sensitivity analysis and simulation validation therefore motivate matching audit coverage and uncertainty to the intended inference \citep{saltelli2008global,sargent2010verification}.
\subsection{Repeated Prisoner's Dilemma as a Goal-Structured Testbed}

Repeated Prisoner's Dilemma provides a controlled testbed for cooperation and adaptation in LLM agents \citep{akata2025repeatedgames,fontana2025nicer}. Fixed actions, payoffs, and horizon let us compare representation sensitivity across fixed-opponent and reciprocal LLM--LLM interaction.

The three perturbations target how the agent, game, and history are represented. \textbf{Persona format} tests whether matched strategic content yields stable behavior across presentation formats, complementing concerns about persona validity \citep{li2025llm}. \textbf{Game framing} tests alternative descriptions of the same payoff structure, motivated by framing sensitivity in strategic settings \citep{lore2024strategic}. \textbf{Memory representation} tests how tabular or narrative histories and summary statistics affect decisions, building on work examining LLMs' use of gameplay logs \citep{fontana2025nicer}. These contrasts separate representation dependence from changes to the game's rules.

\subsection{Echo Chambers as an Open-Ended Social Simulation Testbed}

Echo-chamber simulations provide a complementary setting in which collective structure emerges from local engagement rather than a shared payoff. Prior studies connect network structure, selective exposure, and agent behavior to polarization and echo-chamber formation \citep{gu2025echochamber,wang2025decoding,ferraro2024agent,piao2025emergence}. We use this setting to trace sensitivity from interaction targeting to the network formed by those choices.

We first vary \textbf{persona representation} while preserving persona information, testing whether presentation alone changes targeting and collective structure. The design perturbations then follow the order of the main outcome summary: \textbf{activation probability} changes participation opportunities; \textbf{memory window} changes the recent interaction context; and \textbf{recommendation feed size} changes visible exposure, a central component of echo-chamber models \citep{wang2025decoding,tornberg2023simulating}. Finally, \textbf{input-network homophily} varies initial like-minded connectivity, and \textbf{hub assignment} varies which stance group occupies highly connected positions while preserving the degree sequence. These structural contrasts complement the main analyses and provide context-relevant conditions for the Audit Card, rather than alternative representations of a fixed system.

\newpage
\section{Additional Experimental Details}
\label{appendix:simulation-details}

\subsection{Repeated Prisoner's Dilemma}
\label{appendix:pd-details}

\subsubsection{Setup and Configurations}
\label{appendix:pd-setup}

Each Prisoner's Dilemma episode runs for $T=10$ rounds. In each round, each player simultaneously chooses Cooperate ($C$) or Defect ($D$), with payoffs $(C,C)=(3,3)$, $(C,D)=(0,5)$, $(D,C)=(5,0)$, and $(D,D)=(1,1)$.

We evaluate two interaction regimes. In the \emph{single-agent} regime, one LLM-controlled agent plays separately against four fixed policies: Tit-for-Tat, Random, Always-Cooperate, and Always-Defect. Tit-for-Tat cooperates in the first round and subsequently copies the opponent's previous action; Random selects $C$ or $D$ with equal probability; Always-Cooperate and Always-Defect select the corresponding action in every round. In the \emph{two-agent} regime, two LLM-controlled agents with matched configurations play one another. Following prior work, these fixed policies provide benchmark interaction contexts with distinct response structures \citep{akata2025repeatedgames,fontana2025nicer,di2023recognition}.

We evaluate \texttt{gpt-5.2}, \texttt{claude-haiku-4-5}, \texttt{gemini-2.5-flash}, and \texttt{deepseek-v3.2}. Each condition contains $N=30$ simulation runs. In two-agent runs, both agents are queried once per round and prompt order is randomized across rounds to reduce stable positional effects. If a response cannot be parsed as a valid action after one retry, the simulation falls back to \textsc{Defect}. Decoding and parsing settings are held fixed within each comparison; unless otherwise specified, model calls use temperature $=0.3$ and top-$p=0.95$.

We report condition means and signed mean differences, with 95\% confidence intervals for selected contrasts and interaction-amplification estimates. Pairwise outcome comparisons use two-sided paired tests with Holm correction within each prespecified comparison family \citep{holm1979simple}. For the Prisoner's Dilemma, correction is applied separately within each model, representation dimension, and outcome across its prespecified pairwise contrasts. Interaction-amplification analyses use matched simulation seeds and paired-seed bootstrap procedures as described in Appendix~\ref{appendix:pd-amplification}.

\subsubsection{Outcome Measures}
\label{appendix:pd-metrics}

For agent $i$, let $a_{i,t}\in\{C,D\}$ denote the action selected at round $t$, and let $u_{i,t}$ denote the corresponding payoff.

\paragraph{Average payoff.}
Average payoff is the mean per-round reward:
\[
\operatorname{Payoff}_i
=
\frac{1}{T}
\sum_{t=1}^{T}u_{i,t}.
\]

\paragraph{Cooperation rate.}
Cooperation rate is the fraction of rounds in which the agent selects $C$:
\[
\operatorname{CoopRate}_i
=
\frac{1}{T}
\sum_{t=1}^{T}
\mathbb{I}(a_{i,t}=C).
\]

Cooperation rate is the primary outcome for comparing representation sensitivity across interaction regimes, with average payoff providing the corresponding outcome under the game's incentive structure.

\subsubsection{Perturbation Details}
\label{appendix:pd-perturbations}

We evaluate three representation dimensions while holding the payoff structure and non-targeted prompt components fixed: persona format, game-instruction framing, and memory representation.

\paragraph{Persona format.}
We present a matched strategic persona in three formats: plain prose (\textsc{Plain}), a descriptive bullet list (\textsc{Descriptive}), and a structured key--value table (\textsc{Tabular}). The persona specifies a strategic participant focused on maximizing long-term payoff, studying the opponent for exploitable patterns, and adapting its strategy while remaining competitive.

\paragraph{Plain strategic persona.}

\begin{promptbox}
You are a strategic participant focused on maximizing your long-term payoff. Study opponent behavior for exploitable patterns, and adjust your strategy to gain advantage while staying competitive.
\end{promptbox}

\paragraph{Descriptive strategic persona.}

\begin{promptbox}
\textbf{Persona profile:}\\
\hspace*{1em}- Maximize your long-term payoff\\
\hspace*{1em}- Study opponent for exploitable patterns\\
\hspace*{1em}- Adjust strategy to gain advantage\\
\hspace*{1em}- Stay strategic and competitive
\end{promptbox}

\paragraph{Tabular strategic persona.}

\begin{promptbox}
\texttt{Persona | Value}\\
\texttt{Goal | Maximize your long-term payoff}\\
\texttt{Observation focus | Study opponent for exploitable patterns}\\
\texttt{Strategy | Adjust strategy to gain advantage}\\
\texttt{Style | Stay strategic and competitive}
\end{promptbox}

\paragraph{Game-instruction framing.}
We hold the persona fixed to \textsc{Plain} and vary the framing of the fixed payoff structure. \textsc{Canonical} uses standard game-theoretic labels such as ``Cooperate'' and ``Defect''; \textsc{Moralized} describes the actions using cooperation and exploitation; and \textsc{Risk} describes the same decision in terms of risk, vulnerability, and potential reward.

\paragraph{Canonical game-theory framing.}

\begin{promptbox}
You are playing a repeated Prisoner's Dilemma for 10 rounds.\\
\vspace{0.8em}

\textbf{In each round, you choose one of two actions:}\\
- \texttt{Cooperate}\\
- \texttt{Defect}\\
\vspace{0.8em}

\textbf{The payoffs for each round are:}\\
- If both players cooperate: both receive 3 points.\\
- If you cooperate and the other defects: you receive 0, the other receives 5.\\
- If you defect and the other cooperates: you receive 5, the other receives 0.\\
- If both defect: both receive 1.\\
\vspace{0.8em}

Your goal is to maximize your total points over all rounds.\\
\vspace{0.8em}

At each round, you will see the history of previous rounds.
\end{promptbox}

\paragraph{Moralized framing.}

\begin{promptbox}
You will interact with another participant for 10 rounds.\\
\vspace{0.8em}

\textbf{In each round, you must decide whether to:}\\
- Cooperate fairly with the other person.\\
- Exploit the other person for personal gain.\\
\vspace{0.8em}

\textbf{The outcomes are:}\\
- If both cooperate fairly: both receive 3 points.\\
- If you cooperate but they exploit you: you receive 0, they receive 5.\\
- If you exploit them while they cooperate: you receive 5, they receive 0.\\
- If both exploit each other: both receive 1.\\
\vspace{0.8em}

Your goal is to maximize your total points.\\
\vspace{0.8em}

At each round, you will see past interactions.
\end{promptbox}

\paragraph{Risk framing.}

\begin{promptbox}
You will make decisions over 10 rounds against another participant.\\
\vspace{0.8em}

\textbf{Each round involves a choice:}\\
- Cooperate, which is safer but vulnerable.\\
- Defect, which is riskier but potentially higher reward.\\
\vspace{0.8em}

\textbf{Outcomes:}\\
- Both cooperate: both get 3.\\
- You cooperate, other defects: you get 0, they get 5.\\
- You defect, other cooperates: you get 5, they get 0.\\
- Both defect: both get 1.\\
\vspace{0.8em}

Your goal is to maximize total payoff.\\
\vspace{0.8em}

At each round, you will see previous outcomes.
\end{promptbox}

\paragraph{Memory representation.}
We hold persona and game framing fixed and vary how the interaction history is presented. The design crosses history format (table vs.\ narrative) with the presence or absence of summary statistics, yielding \textsc{Table-Raw}, \textsc{Narrative-Raw}, \textsc{Table+Stats}, and \textsc{Narrative+Stats}. Summary statistics include cooperation rates, cumulative payoffs, and joint outcome counts.

\paragraph{M1: Table, raw history only.}

\begin{promptbox}
\textbf{History:}\\
\texttt{Round | You | Opponent | Your payoff | Opp payoff}\\
\texttt{1 | [YOUR ACTION] | [OPPONENT ACTION] | [YOUR PAYOFF] | [OPPONENT PAYOFF]}\\
\texttt{2 | [YOUR ACTION] | [OPPONENT ACTION] | [YOUR PAYOFF] | [OPPONENT PAYOFF]}\\
\texttt{\ldots}
\end{promptbox}

\paragraph{M2: Narrative, raw history only.}

\begin{promptbox}
\textbf{History:}\\
In round 1, you chose \texttt{[YOUR ACTION]} and the opponent chose \texttt{[OPPONENT ACTION]}. You received \texttt{[YOUR PAYOFF]} points and the opponent received \texttt{[OPPONENT PAYOFF]}.\\
In round 2, you chose \texttt{[YOUR ACTION]} and the opponent chose \texttt{[OPPONENT ACTION]}. You received \texttt{[YOUR PAYOFF]} points and the opponent received \texttt{[OPPONENT PAYOFF]}.\\
\texttt{\ldots}
\end{promptbox}

\paragraph{M3: Table, raw history plus statistics.}

\begin{promptbox}
\textbf{History:}\\
\texttt{Round | You | Opponent | Your payoff | Opp payoff}\\
\texttt{1 | [YOUR ACTION] | [OPPONENT ACTION] | [YOUR PAYOFF] | [OPPONENT PAYOFF]}\\
\texttt{2 | [YOUR ACTION] | [OPPONENT ACTION] | [YOUR PAYOFF] | [OPPONENT PAYOFF]}\\
\texttt{\ldots}\\
\vspace{0.8em}

\textbf{Summary statistics:}\\
- You cooperated \texttt{[N]} times and defected \texttt{[N]} times.\\
- Your cooperation rate: \texttt{[RATE]}.\\
- Opponent cooperated \texttt{[N]} times and defected \texttt{[N]} times.\\
- Opponent cooperation rate: \texttt{[RATE]}.\\
- Your cumulative payoff: \texttt{[PAYOFF]}.\\
- Opponent cumulative payoff: \texttt{[PAYOFF]}.\\
- Your average payoff per round: \texttt{[AVG]}.\\
- Opponent average payoff per round: \texttt{[AVG]}.\\
- Joint outcomes: \texttt{CC=[N], CD=[N], DC=[N], DD=[N]}.
\end{promptbox}

\paragraph{M4: Narrative, raw history plus statistics.}

\begin{promptbox}
\textbf{History:}\\
In round 1, you chose \texttt{[YOUR ACTION]} and the opponent chose \texttt{[OPPONENT ACTION]}. You received \texttt{[YOUR PAYOFF]} points and the opponent received \texttt{[OPPONENT PAYOFF]}.\\
In round 2, you chose \texttt{[YOUR ACTION]} and the opponent chose \texttt{[OPPONENT ACTION]}. You received \texttt{[YOUR PAYOFF]} points and the opponent received \texttt{[OPPONENT PAYOFF]}.\\
\texttt{\ldots}\\
\vspace{0.8em}

\textbf{Summary statistics:}\\
- You cooperated \texttt{[N]} times and defected \texttt{[N]} times.\\
- Your cooperation rate: \texttt{[RATE]}.\\
- Opponent cooperated \texttt{[N]} times and defected \texttt{[N]} times.\\
- Opponent cooperation rate: \texttt{[RATE]}.\\
- Your cumulative payoff: \texttt{[PAYOFF]}.\\
- Opponent cumulative payoff: \texttt{[PAYOFF]}.\\
- Your average payoff per round: \texttt{[AVG]}.\\
- Opponent average payoff per round: \texttt{[AVG]}.\\
- Joint outcomes: \texttt{CC=[N], CD=[N], DC=[N], DD=[N]}.
\end{promptbox}

For the two-agent memory condition, the same four memory variants are used, but the history is role-specific. For Agent A, ``You'' refers to Agent A and ``Other agent'' refers to Agent B; for Agent B, the mapping is reversed.

\subsubsection{Interaction-Amplification Measures}
\label{appendix:pd-amplification}

For model $m$, representation contrast $p=(c_0,c_1)$, and interaction context $o$, we define the signed perturbation effect on cooperation as
\[
\Delta_{m,p,o}
=
\mathbb{E}[Y\mid c_1,m,o]
-
\mathbb{E}[Y\mid c_0,m,o],
\]
where $Y$ denotes cooperation rate. For the fixed-opponent regime,
\[
\mathcal{O}
=
\{\text{Tit-for-Tat},\text{Random},
\text{Always-Cooperate},\text{Always-Defect}\}.
\]

We summarize sensitivity across fixed-opponent contexts as
\[
S^{\mathrm{fixed}}_{m,p}
=
\frac{1}{|\mathcal{O}|}
\sum_{o\in\mathcal{O}}
|\Delta_{m,p,o}|,
\]
and sensitivity under reciprocal LLM--LLM interaction as
\[
S^{\mathrm{LLM}}_{m,p}
=
|\Delta_{m,p,\mathrm{LLM}}|.
\]

The interaction effect gap is
\[
A_{m,p}
=
S^{\mathrm{LLM}}_{m,p}
-
S^{\mathrm{fixed}}_{m,p}.
\]
Positive $A_{m,p}$ indicates \emph{amplification}, while negative $A_{m,p}$ indicates \emph{attenuation}. The fixed-opponent measure averages the absolute opponent-specific effects before comparison with the LLM--LLM effect, preserving sensitivity that differs in direction across benchmark interaction contexts.

We estimate uncertainty in $A_{m,p}$ using $10{,}000$ paired bootstrap replicates over the $30$ matched simulation seeds. Within each replicate, the same sampled seeds are used across representation conditions, fixed opponents, and the LLM--LLM regime. We report pointwise percentile 95\% confidence intervals for $A_{m,p}$ and compute the corresponding quantities by round to characterize temporal variation.

\subsection{Social-Media Echo Chamber}
\label{appendix:ec-details}

\subsubsection{Setup and Configurations}
\label{appendix:ec-setup}

The echo-chamber simulation contains $N=100$ agents interacting for $T=15$ rounds on a social network. Each agent has a persona and a fixed stance $s_i\in\{-2,-1,0,1,2\}$ toward stronger regulation of advanced AI systems. Agent stances remain fixed throughout each run, so the simulation measures interaction structure conditional on the initial stance distribution.

In each round, agents become active according to the configured activation probability. If no agent is sampled as active, one agent is selected at random so that every round contains at least one decision. Active agents observe recent posts from direct network neighbors and choose among \textsc{Post}, \textsc{Repost}, \textsc{Reply}, and \textsc{Do Nothing}. Reposts and replies additionally select a visible interaction target. The default configuration uses a feed size of $5$ posts and a memory window of $2$ rounds. Invalid repost or reply targets are converted according to the simulation's parsing rules while preserving the fixed action space.

We evaluate \texttt{gpt-5.2}, \texttt{claude-haiku-4-5}, \texttt{gemini-2.5-flash}, and \texttt{deepseek-v4-flash}. Unless otherwise specified, each condition contains $N=30$ simulation runs. Decoding and parsing settings are held fixed within each comparison.

\paragraph{Network construction.}
To control overall connectivity across structural conditions, we construct networks from a fixed power-law-like degree sequence with target mean degree $8$ and exponent $2.4$. The sequence is adjusted until graphical, producing a heavy-tailed network in which a small number of agents occupy high-degree positions. The same degree sequence is reused across network variants, holding the degree distribution and overall connectivity fixed.

\subsubsection{Outcome Measures}
\label{appendix:ec-metrics}

We measure representation sensitivity at both the agent and collective levels. For paired persona-representation runs, agent decisions are matched by $(\text{seed},\text{round},\text{agent})$ across the \textsc{Bullet} and \textsc{Prose} conditions.

\paragraph{Action agreement.}
For two paired conditions $c_0$ and $c_1$, action agreement measures whether the same agent selects the same action at the same round. For paired decision $k$,
\[
I^{\mathrm{action}}_k
=
\mathbb{I}
\left(
a_k^{(c_1)}
=
a_k^{(c_0)}
\right).
\]
For seed $s$, action agreement is
\[
G^{\mathrm{action}}_s
=
\frac{1}{|\mathcal{K}_s|}
\sum_{k\in\mathcal{K}_s}
I^{\mathrm{action}}_k,
\]
where $\mathcal{K}_s$ contains the matched agent--round decisions for that seed. We compute agreement within each seed and then summarize across paired seeds.

\paragraph{Target agreement.}
Target agreement measures whether paired decisions select the same non-missing interaction-target author:
\[
I^{\mathrm{target}}_k
=
\mathbb{I}
\left(
t_k^{(c_1)}
=
t_k^{(c_0)}
\neq \varnothing
\right),
\]
where $t_k$ denotes the selected target author. Different targets, a target in only one condition, and no target in either condition are counted as non-agreement. Seed-level target agreement is
\[
G^{\mathrm{target}}_s
=
\frac{1}{|\mathcal{K}_s|}
\sum_{k\in\mathcal{K}_s}
I^{\mathrm{target}}_k.
\]

For the persona-representation analysis, $c_0=\textsc{Bullet}$ and $c_1=\textsc{Prose}$.

\paragraph{Weighted stance assortativity.}
Let $G_I=(V,E_I)$ denote the realized interaction graph, where an edge from agent $i$ to agent $j$ records a repost or reply, and let $A_{ij}$ denote the number of observed interactions from $i$ to $j$. Let $g_i\in\mathcal{G}$ denote agent $i$'s stance group. For stance groups $u,v\in\mathcal{G}$, define the fraction of weighted interactions from group $u$ to group $v$ as
\[
p_{uv}
=
\frac{
\sum_{i,j}A_{ij}\mathbb{I}(g_i=u,g_j=v)
}{
\sum_{i,j}A_{ij}
},
\]
with marginals
\[
p_u^{\mathrm{out}}
=
\sum_{v\in\mathcal{G}}p_{uv},
\qquad
p_u^{\mathrm{in}}
=
\sum_{v\in\mathcal{G}}p_{vu}.
\]
Weighted stance assortativity is
\[
r
=
\frac{
\sum_{u\in\mathcal{G}}p_{uu}
-
\sum_{u\in\mathcal{G}}p_u^{\mathrm{out}}p_u^{\mathrm{in}}
}{
1-
\sum_{u\in\mathcal{G}}p_u^{\mathrm{out}}p_u^{\mathrm{in}}
}.
\]
Positive values indicate greater within-stance interaction than expected from the observed interaction marginals, while negative values indicate greater cross-stance interaction.

\paragraph{Weighted same-group edge ratio.}
We additionally measure the fraction of interaction volume occurring within stance groups:
\[
H
=
\frac{
\sum_{i,j}A_{ij}\mathbb{I}(g_i=g_j)
}{
\sum_{i,j}A_{ij}
}.
\]
Larger values indicate a greater concentration of interaction within stance groups.

\subsubsection{Perturbation Details}
\label{appendix:ec-perturbations}

We evaluate one representation perturbation and five design perturbations. The representation perturbation varies persona presentation while preserving persona information. The design perturbations vary input-network homophily, hub assignment, activation probability, memory-window length, and recommendation feed size.

\paragraph{Persona representation.}
We present matched persona information either as structured bullets (\textsc{Bullet}) or narrative prose (\textsc{Prose}), while holding the remaining simulation configuration fixed.

\paragraph{Structured bullet-point persona (original).}

\begin{promptbox}
Your profile:\\
\hspace*{1em}- agent\_id: \{agent\_id\}\\
\hspace*{1em}- bio: \{bio\}\\
\hspace*{1em}- follower\_count: \{follower\_count\}\\
\hspace*{1em}- current stance on AI regulation: \{stance\}\\
\hspace*{2em}(where $-2$ strongly oppose, $-1$ somewhat oppose,\\
\hspace*{2em}$0$ mixed/undecided, $1$ somewhat support,\\
\hspace*{2em}$2$ strongly support)\\
\hspace*{1em}- neighbor count in the network: \{neighbor\_count\}
\end{promptbox}

\paragraph{Narrative-prose persona.}

\begin{promptbox}
You are \{agent\_id\}, \{bio\}. You have \{follower\_count\} followers, and \{neighbor\_count\} people in the network whose posts you can see. On the question of whether governments should impose stronger regulation on advanced AI systems, you currently \{stance\_description\}.
\end{promptbox}

Here, \texttt{\{stance\_description\}} is the prose equivalent of the numerical stance: strongly oppose, somewhat oppose, hold a mixed or undecided view, somewhat support, or strongly support stronger regulation.

\paragraph{Interaction-level design perturbations.}
Activation probability compares $0.3$ and $0.5$ for all models. Memory window compares $2$ and $4$ rounds, and recommendation feed size compares up to $5$ and $10$ recent neighbor posts.

\paragraph{Input-network homophily (context-relevant design-level perturbation).}
We vary stance homophily while preserving the fixed degree sequence. Let $G_0=(V,E_0)$ denote the input network and $x_i$ agent $i$'s stance on the five-point scale. Starting from a graph realizing the fixed degree sequence, we apply degree-preserving double-edge swaps to obtain three input-network assortativity ranges:
\[
\textsc{Low}:~h(G_0)\in[0.02,0.10],
\qquad
\textsc{Medium}:~h(G_0)\in[0.12,0.20],
\qquad
\textsc{High}:~h(G_0)\in[0.22,0.30].
\]
The rewiring procedure changes which stance groups are connected while preserving each agent's degree.

\paragraph{Hub assignment (context-relevant design-level perturbation).}
We vary which agents occupy high-degree positions while preserving the degree sequence. The \textsc{Random Hub} condition randomly assigns degrees across agents, while \textsc{Mixed Hub} distributes high-degree positions across stance groups. The \textsc{Pro Hub} and \textsc{Anti Hub} conditions preferentially assign high-degree positions to the corresponding stance group. Networks are subsequently rewired to the medium-homophily range, $h(G_0)\in[0.12,0.20]$, approximately holding stance homophily fixed across hub conditions.

\paragraph{Statistical comparisons.}
Conditions are matched by simulation seed. Binary design perturbations are evaluated using two-sided paired $t$-tests. Holm correction is applied separately within each model and outcome across activation probability, memory window, and recommendation feed size. Persona-representation effects are evaluated using two-sided paired Wilcoxon signed-rank tests, with Holm correction across the four models and two collective outcomes. All significance markers in the outcome-sensitivity tables use the corresponding Holm-corrected $p$-values.

\subsubsection{Propagation Measures}
\label{appendix:ec-propagation}

For perturbation $p=(c_0,c_1)$, let $q_t^{(c)}$ denote the cross-group targeting rate under condition $c$ at round $t$. The signed behavioral effect and its sensitivity magnitude are
\[
\Delta_{\mathrm{target},t}(p)
=
q_t^{(c_1)}-q_t^{(c_0)},
\qquad
S_{\mathrm{target},t}(p)
=
\left|\Delta_{\mathrm{target},t}(p)\right|.
\]
For paired persona-representation runs, we additionally compute action-type agreement and interaction-target agreement to locate representation sensitivity in action selection or interaction targeting.

For collective network outcome $Y\in\{r,H\}$, we define
\[
\Delta_{Y,t}(p)
=
Y_t^{(c_1)}-Y_t^{(c_0)},
\qquad
S_{Y,t}(p)
=
\left|\Delta_{Y,t}(p)\right|.
\]
We trace representation sensitivity across interaction levels by comparing the temporal development of behavioral effects and collective network effects.

% \paragraph{Exposure diagnostic.}
% For persona-representation runs, we additionally measure cross-group exposure in the visible feed. Let $e_{i,t}^{(c)}$ denote the fraction of visible feed content shown to agent $i$ at round $t$ that originates from other stance groups. The condition-level exposure difference is
% \[
% \Delta_{\mathrm{exp},t}(p)
% =
% \mathbb{E}_i\!\left[e_{i,t}^{(c_1)}\right]
% -
% \mathbb{E}_i\!\left[e_{i,t}^{(c_0)}\right].
% \]
% This quantity is used as a diagnostic of whether persona representation changes the stance composition of agents' visible information environments.

\newpage
\section{Extended Experimental Results}
\label{appendix:extended-results}

\begin{figure}[t]
    \centering
    \includegraphics[width=\textwidth]{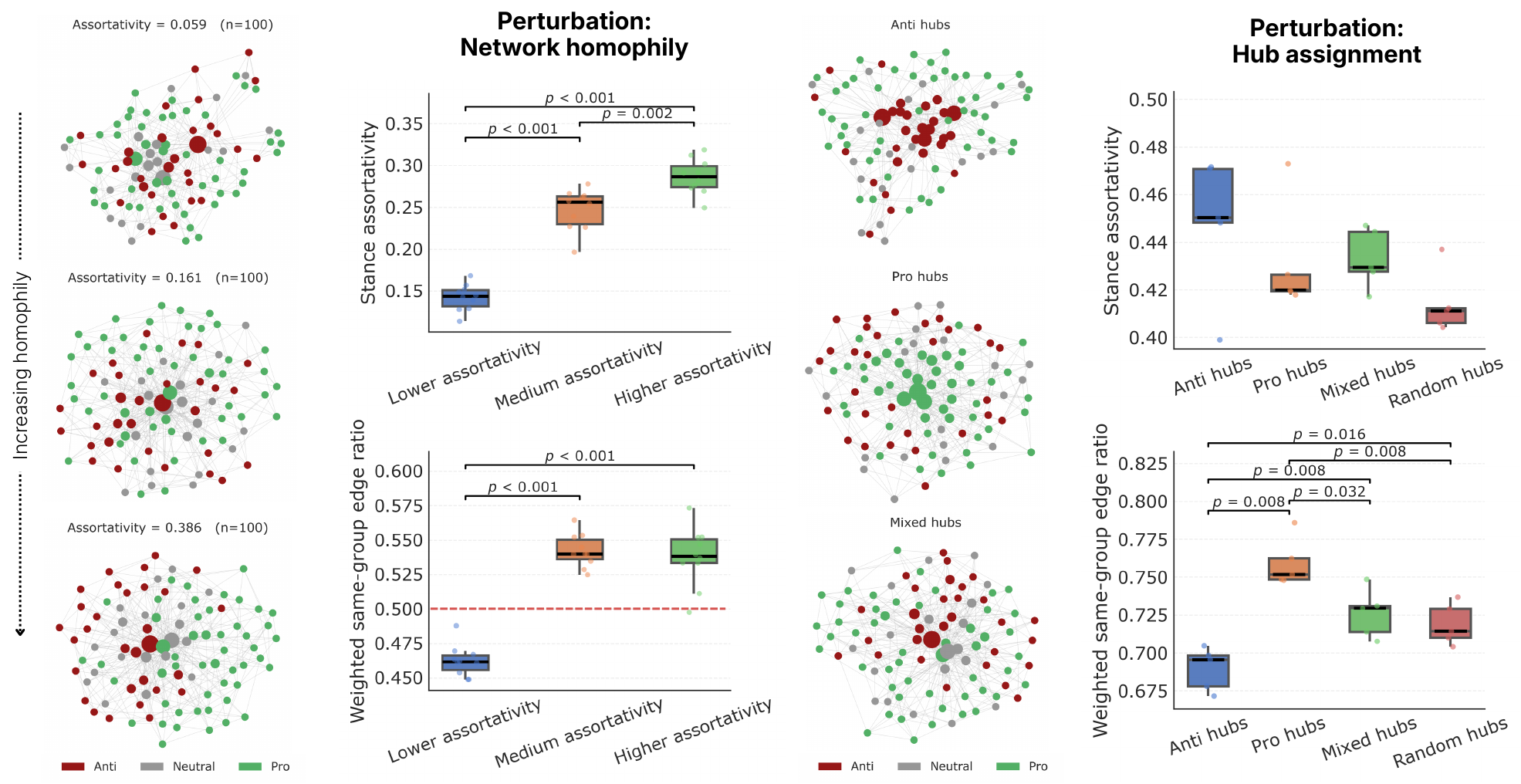}
    \caption{
\textbf{Effects of network homophily and hub assignment on echo-chamber outcomes.}
Left panels vary initial stance homophily; right panels vary hub assignment while holding the degree sequence fixed. Boxplots show run-level stance assortativity and weighted same-group edge ratio. Significant pairwise differences are based on paired t-tests with Holm correction ($p<0.05$).
}
    \label{fig:homophily_hub_panel}
\end{figure}

This section provides extended results for the Prisoner's Dilemma (PD) and echo-chamber experiments. For the PD experiment, Table~\ref{tab:sensitivity-summary-full} reports pairwise outcome contrasts across perturbations, representations, and models, while Table~\ref{tab:appendix-pd-complete-amplification} provides the complete interaction-amplification results for the prespecified representation contrasts. For the echo-chamber experiment, Table~\ref{tab:ec-representation-full} reports representation sensitivity in both behavioral and collective network outcomes. Table~\ref{tab:ec-design-full} reports signed design-perturbation contrasts with confidence intervals across all four models. We additionally report the effects of network homophily and hub assignment in Figure~\ref{fig:homophily_hub_panel}. Although these two perturbations are not analyzed in the main text, they serve as the audit-card examples of context-relevant perturbations; we therefore include their corresponding experimental results here for completeness.

\begin{table*}[t]
\centering
\caption{\textbf{Pairwise contrasts in the Prisoner's Dilemma experiment across perturbations, representations, and models.}
$\Delta$ denotes the difference in outcome means between two perturbation conditions.}
\label{tab:sensitivity-summary-full}
\vspace{0.5em}

\scriptsize
\setlength{\tabcolsep}{4.5pt}
\renewcommand{\arraystretch}{1.12}

\begin{tabular}{
    >{\raggedright\arraybackslash}p{1.5cm}
    >{\raggedright\arraybackslash}p{1.5cm}
    >{\raggedright\arraybackslash}p{2.5cm}
    cccc
}
\toprule
\textbf{Perturbation}
& \textbf{Outcome}
& \textbf{Contrast}
& \textit{GPT-5.2}
& \textit{Claude Haiku 4.5}
& \textit{Gemini 2.5 Flash}
& \textit{DeepSeek V3.2} \\
\midrule

\multicolumn{7}{c}{\textbf{Repeated Prisoner's Dilemma (Two-Agent)}} \\
\midrule

% ==================== Persona: Average payoff ====================

\multirow{3}{=}{Persona}
& \multirow{3}{=}{Average payoff}
& Descriptive $\rightarrow$ Tabular
& \cellcolor{sigthree}+1.467***
& \cellcolor{sigthree}+1.223***
& -0.010
& -0.127 \\

&
&
Plain $\rightarrow$ Descriptive
& \cellcolor{sigthree}-1.478***
& \cellcolor{sigthree}-1.487***
& \cellcolor{sigthree}-0.777***
& \cellcolor{sigone}-0.013* \\

&
&
Plain $\rightarrow$ Tabular
& -0.012
& \cellcolor{sigthree}-0.263***
& \cellcolor{sigthree}-0.787***
& \cellcolor{sigtwo}-0.140** \\

\addlinespace[2pt]

% ==================== Persona: Cooperation ====================

\multirow{3}{=}{Persona}
& \multirow{3}{=}{Cooperation rate (pp)}
& Descriptive $\rightarrow$ Tabular
& \cellcolor{sigthree}+74.000***
& \cellcolor{sigthree}+63.333***
& -0.333
& -6.333 \\

&
&
Plain $\rightarrow$ Descriptive
& \cellcolor{sigthree}-75.167***
& \cellcolor{sigthree}-77.667***
& \cellcolor{sigthree}-35.667***
& \cellcolor{sigone}-1.333* \\

&
&
Plain $\rightarrow$ Tabular
& -1.167
& \cellcolor{sigthree}-14.333***
& \cellcolor{sigthree}-36.000***
& \cellcolor{sigtwo}-7.667** \\

\addlinespace[2pt]

% ==================== Game framing: Average payoff ====================

\multirow{3}{=}{Game framing}
& \multirow{3}{=}{Average payoff}
& Canonical $\rightarrow$ Moralized
& \cellcolor{sigthree}+0.417***
& +0.025
& -0.187
& \cellcolor{sigtwo}+0.028** \\

&
&
Canonical $\rightarrow$ Risk
& +0.110
& \cellcolor{sigone}+0.087*
& -0.142
& \cellcolor{sigtwo}-0.120** \\

&
&
Moralized $\rightarrow$ Risk
& \cellcolor{sigtwo}-0.307**
& +0.062
& +0.045
& \cellcolor{sigthree}-0.148*** \\

\addlinespace[2pt]

% ==================== Game framing: Cooperation ====================

\multirow{3}{=}{Game framing}
& \multirow{3}{=}{Cooperation rate (pp)}
& Canonical $\rightarrow$ Moralized
& \cellcolor{sigthree}+19.333***
& +0.833
& -9.000
& \cellcolor{sigtwo}+2.167** \\

&
&
Canonical $\rightarrow$ Risk
& +4.667
& \cellcolor{sigone}+3.333*
& -7.500
& \cellcolor{sigtwo}-6.667** \\

&
&
Moralized $\rightarrow$ Risk
& \cellcolor{sigone}-14.667*
& +2.500
& +1.500
& \cellcolor{sigthree}-8.833*** \\

\addlinespace[2pt]

% ==================== Memory: Average payoff ====================

\multirow{6}{=}{Memory}
& \multirow{6}{=}{Average payoff}
& NR $\rightarrow$ NS
& \cellcolor{sigone}+0.038*
& +0.043
& +0.092
& +0.000 \\

&
&
NR $\rightarrow$ TS
& \cellcolor{sigthree}+0.088***
& +0.048
& +0.017
& +0.005 \\

&
&
TS $\rightarrow$ NS
& \cellcolor{sigtwo}-0.050**
& -0.005
& +0.075
& -0.005 \\

&
&
TR $\rightarrow$ NS
& \cellcolor{sigthree}-0.063***
& +0.045
& +0.285
& +0.008 \\

&
&
TR $\rightarrow$ NR
& \cellcolor{sigthree}-0.102***
& +0.002
& +0.193
& +0.008 \\

&
&
TR $\rightarrow$ TS
& \cellcolor{sigone}-0.013*
& +0.050
& +0.210
& +0.013 \\

\addlinespace[2pt]

% ==================== Memory: Cooperation ====================

\multirow{6}{=}{Memory}
& \multirow{6}{=}{Cooperation rate (pp)}
& NR $\rightarrow$ NS
& \cellcolor{sigone}+1.833*
& +2.333
& +4.833
& -0.000 \\

&
&
NR $\rightarrow$ TS
& \cellcolor{sigthree}+4.833***
& +2.500
& +1.667
& +0.500 \\

&
&
TS $\rightarrow$ NS
& \cellcolor{sigtwo}-3.000**
& -0.167
& +3.167
& -0.500 \\

&
&
TR $\rightarrow$ NS
& \cellcolor{sigthree}-4.333***
& +2.500
& +13.833
& +0.500 \\

&
&
TR $\rightarrow$ NR
& \cellcolor{sigthree}-6.167***
& +0.167
& +9.000
& +0.500 \\

&
&
TR $\rightarrow$ TS
& \cellcolor{sigone}-1.333*
& +2.667
& +10.667
& +1.000 \\

\bottomrule
\end{tabular}

\vspace{2pt}

\begin{minipage}{0.96\textwidth}
\tiny
\textit{Note:}
Values report estimated differences between the indicated perturbation conditions.
Cooperation effects are reported in percentage points; average payoff effects use their native units.
For memory contrasts, NR = Narrative-Raw, NS = Narrative+Stats,
TR = Table-Raw, and TS = Table+Stats.
Shading denotes Holm-corrected significance:
*** $p<.001$, ** $p<.01$, * $p<.05$;
unshaded cells are not statistically significant.
\end{minipage}

\end{table*}

\newcommand{\CI}[2]{\mbox{[#1, #2]}}
\begingroup

% Keep the caption at normal font size while keeping the table body small.
\captionsetup{font=normalsize}

\scriptsize
\setlength{\tabcolsep}{3pt}
\renewcommand{\arraystretch}{1.08}

\begin{longtable}{
    >{\raggedright\arraybackslash}p{2.6cm}
    >{\centering\arraybackslash}p{2.35cm}
    >{\centering\arraybackslash}p{2.35cm}
    >{\centering\arraybackslash}p{2.35cm}
    >{\centering\arraybackslash}p{2.35cm}
}

\caption{\textbf{Interaction amplification across prespecified representation contrasts in the Prisoner's Dilemma experiment.}}
\label{tab:appendix-pd-complete-amplification}
\\

\toprule
\textbf{Contrast}
& \textit{GPT-5.2}
& \textit{Claude Haiku 4.5}
& \textit{Gemini 2.5 Flash}
& \textit{DeepSeek V3.2} \\
\midrule
\endfirsthead

\multicolumn{5}{l}{\scriptsize\textit{Table \thetable\ continued from previous page}}\\
\toprule
\textbf{Contrast}
& \textit{GPT-5.2}
& \textit{Claude Haiku 4.5}
& \textit{Gemini 2.5 Flash}
& \textit{DeepSeek V3.2} \\
\midrule
\endhead

\midrule
\multicolumn{5}{r}{\scriptsize\textit{Continued on next page}}\\
\endfoot

\bottomrule
\endlastfoot

% ============================================================
% Persona format
% ============================================================

\multicolumn{5}{l}{\textbf{Persona format}}\\
\addlinespace[2pt]

Plain vs Descriptive
&
\shortstack{
$S_{\mathrm{fixed}}$ = 0.271\\
$S_{\mathrm{LLM}}$ = 0.752\\
$A$ = +0.481\\
\CI{+0.327}{+0.622}\\
Amplification
}
&
\shortstack{
$S_{\mathrm{fixed}}$ = 0.315\\
$S_{\mathrm{LLM}}$ = 0.777\\
$A$ = +0.462\\
\CI{+0.376}{+0.529}\\
Amplification
}
&
\shortstack{
$S_{\mathrm{fixed}}$ = 0.070\\
$S_{\mathrm{LLM}}$ = 0.357\\
$A$ = +0.287\\
\CI{+0.190}{+0.387}\\
Amplification
}
&
\shortstack{
$S_{\mathrm{fixed}}$ = 0.101\\
$S_{\mathrm{LLM}}$ = 0.013\\
$A$ = -0.087\\
\CI{-0.108}{-0.066}\\
Attenuation
}
\\
\addlinespace[4pt]

Plain vs Tabular
&
\shortstack{
$S_{\mathrm{fixed}}$ = 0.589\\
$S_{\mathrm{LLM}}$ = 0.012\\
$A$ = -0.578\\
\CI{-0.602}{-0.556}\\
Attenuation
}
&
\shortstack{
$S_{\mathrm{fixed}}$ = 0.429\\
$S_{\mathrm{LLM}}$ = 0.143\\
$A$ = -0.286\\
\CI{-0.341}{-0.223}\\
Attenuation
}
&
\shortstack{
$S_{\mathrm{fixed}}$ = 0.074\\
$S_{\mathrm{LLM}}$ = 0.360\\
$A$ = +0.286\\
\CI{+0.192}{+0.384}\\
Amplification
}
&
\shortstack{
$S_{\mathrm{fixed}}$ = 0.055\\
$S_{\mathrm{LLM}}$ = 0.077\\
$A$ = +0.022\\
\CI{-0.050}{+0.121}\\
No detectable difference
}
\\
\addlinespace[4pt]

Descriptive vs Tabular
&
\shortstack{
$S_{\mathrm{fixed}}$ = 0.318\\
$S_{\mathrm{LLM}}$ = 0.740\\
$A$ = +0.422\\
\CI{+0.295}{+0.546}\\
Amplification
}
&
\shortstack{
$S_{\mathrm{fixed}}$ = 0.244\\
$S_{\mathrm{LLM}}$ = 0.633\\
$A$ = +0.389\\
\CI{+0.291}{+0.481}\\
Amplification
}
&
\shortstack{
$S_{\mathrm{fixed}}$ = 0.004\\
$S_{\mathrm{LLM}}$ = 0.003\\
$A$ = -0.001\\
\CI{-0.006}{+0.013}\\
No detectable difference
}
&
\shortstack{
$S_{\mathrm{fixed}}$ = 0.054\\
$S_{\mathrm{LLM}}$ = 0.063\\
$A$ = +0.009\\
\CI{-0.057}{+0.102}\\
No detectable difference
}
\\

\addlinespace[5pt]

% ============================================================
% Game framing
% ============================================================

\multicolumn{5}{l}{\textbf{Game framing}}\\
\addlinespace[2pt]

Canonical vs Moralized
&
\shortstack{
$S_{\mathrm{fixed}}$ = 0.117\\
$S_{\mathrm{LLM}}$ = 0.193\\
$A$ = +0.077\\
\CI{+0.024}{+0.127}\\
Amplification
}
&
\shortstack{
$S_{\mathrm{fixed}}$ = 0.058\\
$S_{\mathrm{LLM}}$ = 0.008\\
$A$ = -0.050\\
\CI{-0.067}{-0.007}\\
Attenuation
}
&
\shortstack{
$S_{\mathrm{fixed}}$ = 0.025\\
$S_{\mathrm{LLM}}$ = 0.090\\
$A$ = +0.065\\
\CI{-0.022}{+0.143}\\
No detectable difference
}
&
\shortstack{
$S_{\mathrm{fixed}}$ = 0.016\\
$S_{\mathrm{LLM}}$ = 0.022\\
$A$ = +0.006\\
\CI{-0.021}{+0.017}\\
No detectable difference
}
\\
\addlinespace[4pt]

Canonical vs Risk
&
\shortstack{
$S_{\mathrm{fixed}}$ = 0.127\\
$S_{\mathrm{LLM}}$ = 0.047\\
$A$ = -0.080\\
\CI{-0.131}{+0.001}\\
No detectable difference
}
&
\shortstack{
$S_{\mathrm{fixed}}$ = 0.086\\
$S_{\mathrm{LLM}}$ = 0.033\\
$A$ = -0.052\\
\CI{-0.088}{-0.017}\\
Attenuation
}
&
\shortstack{
$S_{\mathrm{fixed}}$ = 0.024\\
$S_{\mathrm{LLM}}$ = 0.075\\
$A$ = +0.051\\
\CI{-0.021}{+0.116}\\
No detectable difference
}
&
\shortstack{
$S_{\mathrm{fixed}}$ = 0.072\\
$S_{\mathrm{LLM}}$ = 0.067\\
$A$ = -0.005\\
\CI{-0.058}{+0.050}\\
No detectable difference
}
\\
\addlinespace[4pt]

Moralized vs Risk
&
\shortstack{
$S_{\mathrm{fixed}}$ = 0.243\\
$S_{\mathrm{LLM}}$ = 0.147\\
$A$ = -0.097\\
\CI{-0.186}{-0.012}\\
Attenuation
}
&
\shortstack{
$S_{\mathrm{fixed}}$ = 0.112\\
$S_{\mathrm{LLM}}$ = 0.025\\
$A$ = -0.087\\
\CI{-0.120}{-0.040}\\
Attenuation
}
&
\shortstack{
$S_{\mathrm{fixed}}$ = 0.022\\
$S_{\mathrm{LLM}}$ = 0.015\\
$A$ = -0.007\\
\CI{-0.035}{+0.057}\\
No detectable difference
}
&
\shortstack{
$S_{\mathrm{fixed}}$ = 0.087\\
$S_{\mathrm{LLM}}$ = 0.088\\
$A$ = +0.001\\
\CI{-0.045}{+0.053}\\
No detectable difference
}
\\

\addlinespace[5pt]

% ============================================================
% Memory representation
% ============================================================

\multicolumn{5}{l}{\textbf{Memory representation}}\\
\addlinespace[2pt]

Table-Raw vs Narrative-Raw
&
\shortstack{
$S_{\mathrm{fixed}}$ = 0.046\\
$S_{\mathrm{LLM}}$ = 0.062\\
$A$ = +0.016\\
\CI{-0.002}{+0.029}\\
No detectable difference
}
&
\shortstack{
$S_{\mathrm{fixed}}$ = 0.025\\
$S_{\mathrm{LLM}}$ = 0.002\\
$A$ = -0.023\\
\CI{-0.043}{-0.011}\\
Attenuation
}
&
\shortstack{
$S_{\mathrm{fixed}}$ = 0.027\\
$S_{\mathrm{LLM}}$ = 0.090\\
$A$ = +0.063\\
\CI{-0.001}{+0.128}\\
No detectable difference
}
&
\shortstack{
$S_{\mathrm{fixed}}$ = 0.019\\
$S_{\mathrm{LLM}}$ = 0.005\\
$A$ = -0.014\\
\CI{-0.028}{+0.004}\\
No detectable difference
}
\\
\addlinespace[4pt]

Table-Raw vs Table+Stats
&
\shortstack{
$S_{\mathrm{fixed}}$ = 0.021\\
$S_{\mathrm{LLM}}$ = 0.013\\
$A$ = -0.007\\
\CI{-0.025}{+0.008}\\
No detectable difference
}
&
\shortstack{
$S_{\mathrm{fixed}}$ = 0.048\\
$S_{\mathrm{LLM}}$ = 0.027\\
$A$ = -0.021\\
\CI{-0.068}{-0.003}\\
Attenuation
}
&
\shortstack{
$S_{\mathrm{fixed}}$ = 0.025\\
$S_{\mathrm{LLM}}$ = 0.107\\
$A$ = +0.082\\
\CI{+0.012}{+0.147}\\
Amplification
}
&
\shortstack{
$S_{\mathrm{fixed}}$ = 0.010\\
$S_{\mathrm{LLM}}$ = 0.010\\
$A$ = +0.000\\
\CI{-0.017}{+0.016}\\
No detectable difference
}
\\
\addlinespace[4pt]

Table-Raw vs Narrative+Stats
&
\shortstack{
$S_{\mathrm{fixed}}$ = 0.030\\
$S_{\mathrm{LLM}}$ = 0.043\\
$A$ = +0.013\\
\CI{-0.007}{+0.030}\\
No detectable difference
}
&
\shortstack{
$S_{\mathrm{fixed}}$ = 0.089\\
$S_{\mathrm{LLM}}$ = 0.025\\
$A$ = -0.064\\
\CI{-0.088}{-0.043}\\
Attenuation
}
&
\shortstack{
$S_{\mathrm{fixed}}$ = 0.018\\
$S_{\mathrm{LLM}}$ = 0.138\\
$A$ = +0.121\\
\CI{+0.032}{+0.200}\\
Amplification
}
&
\shortstack{
$S_{\mathrm{fixed}}$ = 0.012\\
$S_{\mathrm{LLM}}$ = 0.005\\
$A$ = -0.007\\
\CI{-0.018}{+0.010}\\
No detectable difference
}
\\
\addlinespace[4pt]

Narrative-Raw vs Table+Stats
&
\shortstack{
$S_{\mathrm{fixed}}$ = 0.052\\
$S_{\mathrm{LLM}}$ = 0.048\\
$A$ = -0.003\\
\CI{-0.021}{+0.013}\\
No detectable difference
}
&
\shortstack{
$S_{\mathrm{fixed}}$ = 0.064\\
$S_{\mathrm{LLM}}$ = 0.025\\
$A$ = -0.039\\
\CI{-0.082}{-0.013}\\
Attenuation
}
&
\shortstack{
$S_{\mathrm{fixed}}$ = 0.023\\
$S_{\mathrm{LLM}}$ = 0.017\\
$A$ = -0.007\\
\CI{-0.033}{+0.080}\\
No detectable difference
}
&
\shortstack{
$S_{\mathrm{fixed}}$ = 0.016\\
$S_{\mathrm{LLM}}$ = 0.005\\
$A$ = -0.011\\
\CI{-0.022}{-0.005}\\
Attenuation
}
\\
\addlinespace[4pt]

Narrative-Raw vs Narrative+Stats
&
\shortstack{
$S_{\mathrm{fixed}}$ = 0.037\\
$S_{\mathrm{LLM}}$ = 0.018\\
$A$ = -0.019\\
\CI{-0.042}{+0.004}\\
No detectable difference
}
&
\shortstack{
$S_{\mathrm{fixed}}$ = 0.114\\
$S_{\mathrm{LLM}}$ = 0.023\\
$A$ = -0.091\\
\CI{-0.113}{-0.068}\\
Attenuation
}
&
\shortstack{
$S_{\mathrm{fixed}}$ = 0.021\\
$S_{\mathrm{LLM}}$ = 0.048\\
$A$ = +0.028\\
\CI{-0.027}{+0.134}\\
No detectable difference
}
&
\shortstack{
$S_{\mathrm{fixed}}$ = 0.016\\
$S_{\mathrm{LLM}}$ = 0.000\\
$A$ = -0.016\\
\CI{-0.023}{-0.002}\\
Attenuation
}
\\
\addlinespace[4pt]

Table+Stats vs Narrative+Stats
&
\shortstack{
$S_{\mathrm{fixed}}$ = 0.014\\
$S_{\mathrm{LLM}}$ = 0.030\\
$A$ = +0.016\\
\CI{-0.009}{+0.028}\\
No detectable difference
}
&
\shortstack{
$S_{\mathrm{fixed}}$ = 0.050\\
$S_{\mathrm{LLM}}$ = 0.002\\
$A$ = -0.048\\
\CI{-0.081}{-0.025}\\
Attenuation
}
&
\shortstack{
$S_{\mathrm{fixed}}$ = 0.011\\
$S_{\mathrm{LLM}}$ = 0.032\\
$A$ = +0.021\\
\CI{-0.023}{+0.099}\\
No detectable difference
}
&
\shortstack{
$S_{\mathrm{fixed}}$ = 0.005\\
$S_{\mathrm{LLM}}$ = 0.005\\
$A$ = -0.000\\
\CI{-0.012}{+0.007}\\
No detectable difference
}
\\

\end{longtable}

\vspace{-0.5em}

\noindent
\begin{minipage}{\textwidth}
\scriptsize
\textit{Note.}
Each model cell reports $S_{\mathrm{fixed}}$, $S_{\mathrm{LLM}}$, $A$ [95\% CI], and the CI-based pattern.
$S_{\mathrm{fixed}}$ is the mean across four fixed opponents of the absolute opponent-specific cooperation-rate effect;
$S_{\mathrm{LLM}}$ is the absolute effect under reciprocal LLM--LLM interaction;
and $A=S_{\mathrm{LLM}}-S_{\mathrm{fixed}}$.
CIs use 30 paired seeds and 10,000 replicates; they are pointwise and not adjusted for multiple comparisons.
\end{minipage}

\endgroup

\begin{table*}[t]
\centering
\caption{\textbf{Representation sensitivity in the echo-chamber simulation.}
Results compare \textsc{Prose} with \textsc{Bullet} persona representations across $N=30$ paired seeds. Differences are \textsc{Prose}-\textsc{Bullet}.}
\label{tab:ec-representation-full}

\scriptsize
\setlength{\tabcolsep}{4pt}
\renewcommand{\arraystretch}{1.15}

\begin{tabular}{lcccc}
\toprule
& \textit{GPT-5.2}
& \textit{Claude Haiku 4.5}
& \textit{Gemini 2.5 Flash}
& \textit{DeepSeek V4 Flash} \\
\midrule

\multicolumn{5}{l}{\textit{Paired behavioral agreement}} \\

Action agreement
& \makecell{0.997\\{[0.996, 0.998]}}
& \makecell{0.936\\{[0.932, 0.941]}}
& \makecell{0.850\\{[0.839, 0.860]}}
& \makecell{0.663\\{[0.649, 0.677]}} \\

Target agreement
& \makecell{0.551\\{[0.541, 0.560]}}
& \makecell{0.593\\{[0.585, 0.601]}}
& \makecell{0.610\\{[0.602, 0.618]}}
& \makecell{0.144\\{[0.137, 0.151]}} \\

\addlinespace[2pt]
\multicolumn{5}{l}{\textit{Behavioral sensitivity}} \\

Cross-group targeting (pp)
& \makecell{+5.15\textsuperscript{***}\\{[4.22, 6.07]}}
& \makecell{-0.93\textsuperscript{*}\\{[-1.62, -0.24]}}
& \makecell{+2.31\textsuperscript{***}\\{[1.49, 3.14]}}
& \makecell{-7.35\textsuperscript{***}\\{[-8.98, -5.72]}} \\

\addlinespace[2pt]
\multicolumn{5}{l}{\textit{Collective network sensitivity at round 15}} \\

Stance assortativity
& \makecell{-0.194\textsuperscript{***}\\{[-0.220, -0.167]}}
& \makecell{-0.002\\{[-0.015, 0.010]}}
& \makecell{-0.035\textsuperscript{***}\\{[-0.050, -0.019]}}
& \makecell{-0.023\\{[-0.068, 0.022]}} \\

Same-group edge ratio
& \makecell{-0.051\textsuperscript{***}\\{[-0.061, -0.041]}}
& \makecell{+0.010\textsuperscript{*}\\{[0.003, 0.017]}}
& \makecell{-0.023\textsuperscript{***}\\{[-0.031, -0.014]}}
& \makecell{+0.072\textsuperscript{***}\\{[0.057, 0.087]}} \\

\bottomrule
\end{tabular}

\vspace{2pt}
\begin{minipage}{0.97\textwidth}
\tiny
\textit{Note:} Brackets report 95\% confidence intervals. Cross-group targeting is averaged over rounds $2$--$15$ and reported in percentage points; collective network outcomes are cumulative through round $15$. Significance markers use Holm-corrected paired Wilcoxon $p$-values: $^{***}p<.001$, $^{**}p<.01$, $^{*}p<.05$. Agreement measures are descriptive paired-run measures and are reported with 95\% confidence intervals.
\end{minipage}
\end{table*}

\begin{table*}[t]
\centering
\caption{\textbf{Design sensitivity in the echo-chamber simulation.}
Signed differences compare the higher condition with the lower condition across $N=30$ paired seeds. Outcomes are cumulative through round $15$.}
\label{tab:ec-design-full}
\scriptsize
\setlength{\tabcolsep}{4pt}
\renewcommand{\arraystretch}{1.15}
\begin{tabular}{lcccc}
\toprule
\textbf{Metric}
& \textit{GPT-5.2}
& \textit{Claude Haiku 4.5}
& \textit{Gemini 2.5 Flash}
& \textit{DeepSeek V4 Flash} \\
\midrule
\multicolumn{5}{l}{\textit{Activation probability: $0.5-0.3$}} \\
Stance assortativity
& \makecell{+0.068\textsuperscript{***}\\{[0.041, 0.095]}}
& \makecell{+0.103\textsuperscript{***}\\{[0.085, 0.121]}}
& \makecell{+0.108\textsuperscript{***}\\{[0.089, 0.127]}}
& \makecell{+0.107\textsuperscript{***}\\{[0.080, 0.134]}} \\
Same-group edge ratio
& \makecell{+0.032\textsuperscript{***}\\{[0.019, 0.045]}}
& \makecell{+0.060\textsuperscript{***}\\{[0.048, 0.072]}}
& \makecell{+0.061\textsuperscript{***}\\{[0.050, 0.073]}}
& \makecell{+0.070\textsuperscript{***}\\{[0.051, 0.089]}} \\
\midrule
\multicolumn{5}{l}{\textit{Memory window: $4-2$ rounds}} \\
Stance assortativity
& \makecell{-0.009\\{[-0.035, 0.018]}}
& \makecell{+0.063\textsuperscript{***}\\{[0.049, 0.076]}}
& \makecell{+0.080\textsuperscript{***}\\{[0.065, 0.094]}}
& \makecell{+0.092\textsuperscript{***}\\{[0.057, 0.127]}} \\
Same-group edge ratio
& \makecell{-0.011\\{[-0.021, -0.001]}}
& \makecell{+0.039\textsuperscript{***}\\{[0.031, 0.048]}}
& \makecell{+0.038\textsuperscript{***}\\{[0.028, 0.048]}}
& \makecell{+0.058\textsuperscript{***}\\{[0.040, 0.075]}} \\
\midrule
\multicolumn{5}{l}{\textit{Recommendation feed size: $10-5$ posts}} \\
Stance assortativity
& \makecell{+0.001\\{[-0.018, 0.021]}}
& \makecell{+0.010\\{[-0.002, 0.021]}}
& \makecell{+0.015\textsuperscript{*}\\{[0.003, 0.026]}}
& \makecell{+0.008\\{[-0.024, 0.041]}} \\
Same-group edge ratio
& \makecell{+0.006\\{[-0.003, 0.015]}}
& \makecell{+0.012\textsuperscript{**}\\{[0.005, 0.018]}}
& \makecell{+0.009\textsuperscript{**}\\{[0.002, 0.016]}}
& \makecell{+0.006\\{[-0.009, 0.020]}} \\
\bottomrule
\end{tabular}
\vspace{2pt}
\begin{minipage}{0.97\textwidth}
\tiny
\textit{Note:} Brackets report unadjusted 95\% confidence intervals for paired mean differences. Stars follow the multiplicity-adjusted paired tests described in Appendix~\ref{appendix:ec-details}: $^{***}p<.001$, $^{**}p<.01$, $^{*}p<.05$. An interval excluding zero need not imply significance after adjustment. Absolute differences correspond to Table~\ref{tab:sensitivity-summary}.
\end{minipage}
\end{table*}

\begin{figure*}[t]
\centering
\begin{minipage}[t]{0.32\textwidth}
\centering\scriptsize
\textbf{(a) Activation probability}\\[2pt]
\includegraphics[width=\linewidth]{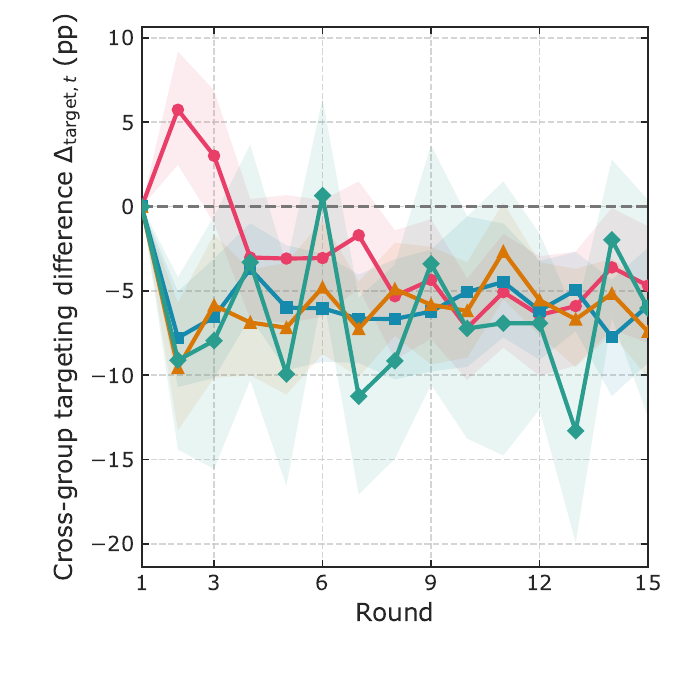}
\end{minipage}\hfill
\begin{minipage}[t]{0.32\textwidth}
\centering\scriptsize
\textbf{(b) Memory window}\\[2pt]
\includegraphics[width=\linewidth]{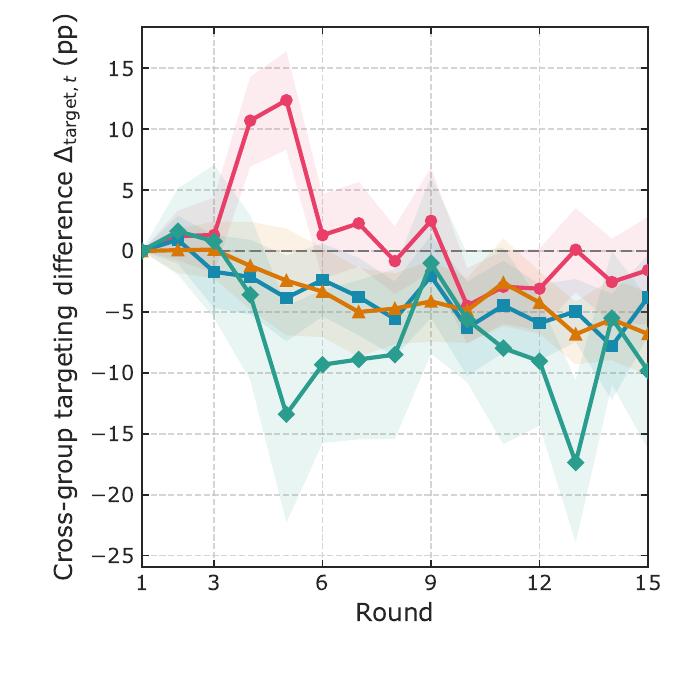}
\end{minipage}\hfill
\begin{minipage}[t]{0.32\textwidth}
\centering\scriptsize
\textbf{(c) Recommendation feed size}\\[2pt]
\includegraphics[width=\linewidth]{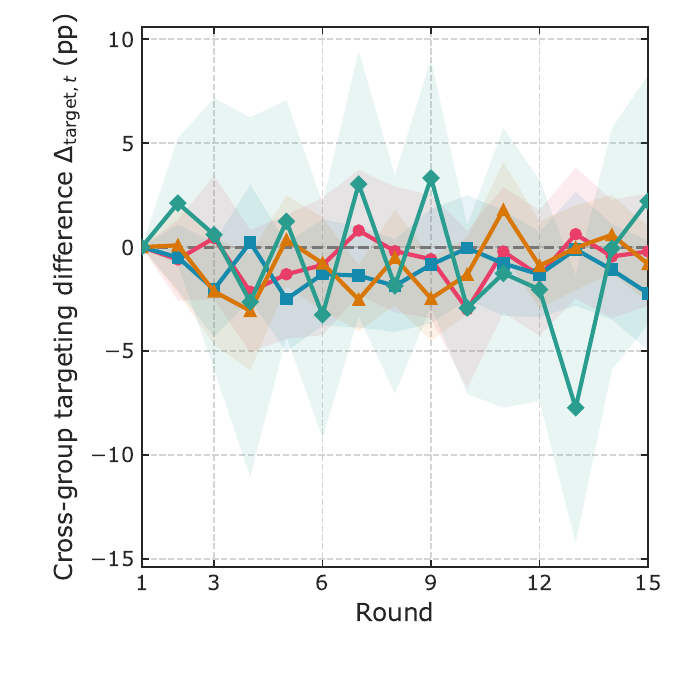}
\end{minipage}
\caption{\textbf{Design sensitivity in interaction targeting.}
Panels report round-wise cross-group targeting differences for (a) activation probability ($0.5-0.3$), (b) memory window ($4-2$ rounds), and (c) recommendation feed size ($10-5$ posts), across four models and $N=30$ paired seeds. Models are GPT-5.2 (pink circles), Claude Haiku 4.5 (blue squares), Gemini 2.5 Flash (orange triangles), and DeepSeek V4 Flash (teal diamonds). Differences are higher minus lower condition, in percentage points; negative values indicate less cross-group targeting under the higher condition. Targeting is measured within each round, not cumulatively. Shaded bands are pointwise 95\% paired-percentile-bootstrap confidence intervals from 10,000 seed resamples, not simultaneous or multiplicity-adjusted intervals.}
\label{fig:ec-design-targeting}
\end{figure*}

\begin{figure*}[t]
\centering
\begin{minipage}[t]{0.32\textwidth}
\centering\scriptsize
\textbf{(a) Activation probability}\\[2pt]
\includegraphics[width=\linewidth]{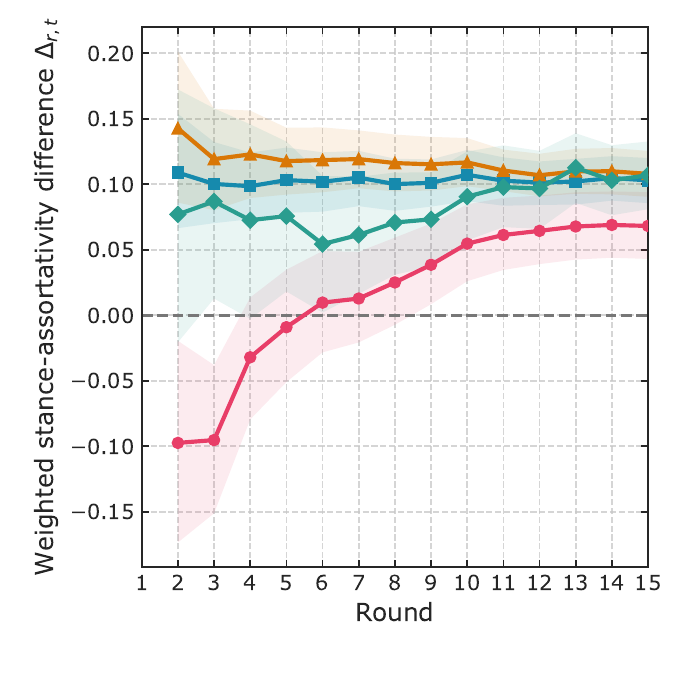}
\end{minipage}\hfill
\begin{minipage}[t]{0.32\textwidth}
\centering\scriptsize
\textbf{(b) Memory window}\\[2pt]
\includegraphics[width=\linewidth]{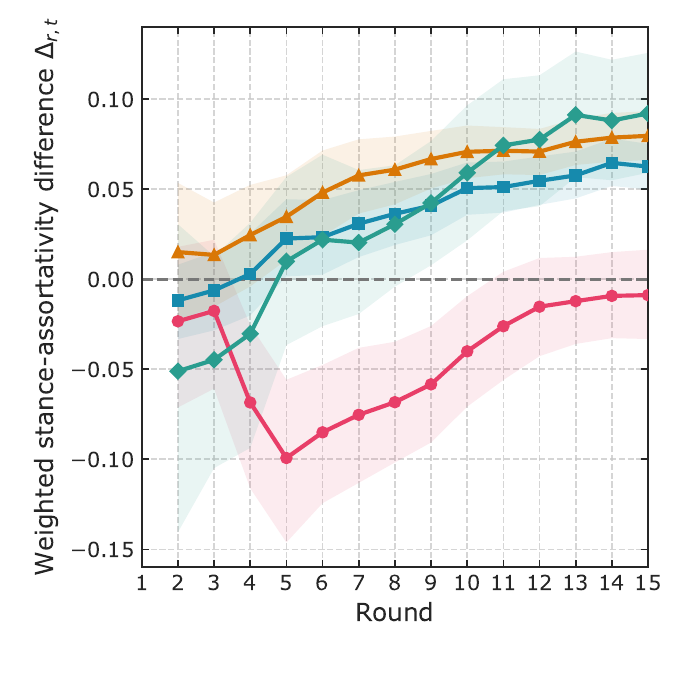}
\end{minipage}\hfill
\begin{minipage}[t]{0.32\textwidth}
\centering\scriptsize
\textbf{(c) Recommendation feed size}\\[2pt]
\includegraphics[width=\linewidth]{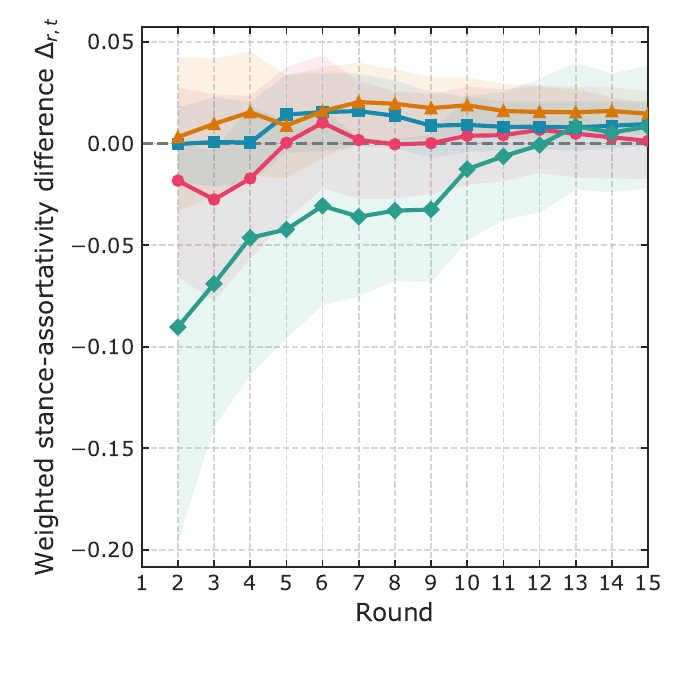}
\end{minipage}
\caption{\textbf{Design sensitivity in collective interaction structure.}
Panels use the same contrasts, models, and paired seeds as Figure~\ref{fig:ec-design-targeting}, and report signed differences in weighted stance assortativity computed from interaction edges accumulated through each round. Positive values indicate higher assortativity under the higher condition. Bands are pointwise 95\% paired-percentile-bootstrap confidence intervals from 10,000 seed resamples. Round 1 is omitted because assortativity is undefined without eligible interaction edges. These trajectories describe collective sensitivity rather than a causal mediation or interaction-amplification estimate.}
\label{fig:ec-design-collective}
\end{figure*}

Figure~\ref{fig:homophily_hub_panel} shows that collective interaction structure depends on initial network homophily and on which stance group occupies hub positions, even when hub reassignment preserves the degree sequence. These context-relevant dependencies qualify the structural conditions under which an echo-chamber claim holds, rather than indicating sensitivity to representation alone.

\section{The Social Simulation Audit Card}
\label{appendix:audit-card}

% \subsection{Which Simulations Need Stronger Robustness Analysis?}

% 1. Identify the simulation scenario and complexity of the simulation based on previous discussion on simulation complexicity ~ robustness strength

% Classify simulation on collective social behavior into 6 types (3 x 2) 
% Goal-structured vs Open ended (theory-drive vs emergent)
% low-stake vs high-stake

% 2. Thinking about the claim type or usage of the simulation results 
% Exploratory probe
% Mechanism evidence
% Policy evidence

TRAILS identifies what can be varied in an audit. Interpreting these perturbations requires an additional, claim-specific judgment: \textit{context relevance}. A dimension is \textit{context-relevant} when it has a meaningful human or theoretical counterpart for the modeled phenomenon, as supported by human studies or established social-science theory. Sensitivity along this dimension identifies conditions under which the claim holds. In contrast, sensitivity along a context-irrelevant dimension reveals implementation dependence. The required audit strength is also shaped by the claim’s evidentiary role, the simulation’s openness, and the stakes of its intended use (see Appendix~\ref{appendix:extended-related-works}). We introduce the \textbf{Social Simulation Audit Card} to operationalize TRAILS for a stated scientific claim. The following five steps connect these requirements to an auditable claim.

\textbf{State the claim.}
Record the scientific conclusion the simulation supports, its evidentiary role, and its intended scope of generalization. Define the property that must be preserved, such as the presence of a phenomenon, an effect direction, or a quantitative effect within a stated tolerance.

\textbf{Map TRAILS and context relevance.}
For each perturbation, identify whether it changes the simulation design (TRAILS-D) or its representation (TRAILS-R), and whether the changed dimension is context-relevant or context-irrelevant. Explain each assignment and record any uncertainty.

\textbf{Choose and register perturbations.} Select perturbations spanning a plausible range for the claim. For each, register the baseline and alternatives, factors held constant, models, and number of independent runs. Identify unaudited dimensions and define the claim-preservation criterion, tolerance, and uncertainty requirements before examining results.

\textbf{Report sensitivity results.} Report effect sizes, uncertainty, and whether the conclusion meets the claim-preservation criterion. Sensitivity to context-relevant perturbations may reflect meaningful social variation. Report the range of conditions under which the claim holds and use this range to define its scope. Substantial changes or reversals under context-irrelevant perturbations indicate implementation dependence. Interpret these results against the specified claim-preservation criterion; statistical significance alone does not determine robustness.

\textbf{Scope the claim according to audit outcomes.}
Classify the claim into one of four outcomes. A claim is \textit{Robust} when it meets the preservation criterion across all audited perturbations with adequate coverage and precision; \textit{Scope-limited} when it holds only within part of the context-relevant conditions tested; \textit{Implementation-sensitive} when context-irrelevant choices, such as persona formatting or memory encoding, substantially change the conclusion; or \textit{Inconclusive} when coverage or precision is insufficient to assess robustness. Report the range supporting a scope-limited claim and bound every verdict by the audited dimensions, configurations, models, and interaction horizons.

Audit sufficiency therefore depends on coverage of the dimensions consequential to the claim and the precision required by its decision rule. Every verdict should state the tested range, model coverage, and remaining uncertainty. Generalization to human behavior additionally requires external validation.

\subsection{Auditing Scientific Claims in the Two Case Studies}
\label{appendix:case-study-audits}

We applied the Social Simulation Audit Card to claims about cooperation and echo-chamber formation. Both experiments used 30 independent runs per condition, with shared seeds for paired comparisons. The audits below summarize the GPT-5.2 results; their verdicts are specific to the reported model, interaction conditions, and outcomes. Table~\ref{tab:case-study-perturbations} maps the reported perturbations to their audit interpretations.

\paragraph{Repeated Prisoner's Dilemma.} We audited the claim that agents exhibited stable cooperation in the two-agent, 10-round repeated Prisoner's Dilemma. Changing the matched persona from \textsc{Plain} to \textsc{Descriptive} reduced cooperation from $98.2\%$ to $23.0\%$, a decrease of $75.2$ percentage points (95\% CI $[62.0,88.3]$), and reduced average payoff by $1.48$ points per round. Because persona format was a context-irrelevant TRAILS-R perturbation, this result supported an \textit{Implementation-sensitive} verdict. Game framing was a context-relevant TRAILS-R perturbation with a maximum cooperation difference of $19.3$ percentage points, indicating that framing conditions should qualify the claim. Memory representation was context-irrelevant and produced a smaller but significant maximum difference of $6.2$ percentage points; whether this difference preserved the claim depends on the specified tolerance. These findings bound the verdict to the audited interaction regime and representation choices.

\begin{table*}[t]
\centering
\scriptsize
\setlength{\tabcolsep}{4pt}
\caption{Audited perturbations and claim-level interpretations for GPT-5.2. PD reports the maximum absolute cooperation-rate difference across contrasts, in percentage points (pp). EC reports absolute differences in stance assortativity ($r$) and weighted same-group edge ratio ($H$). Stars indicate $p<.001$; unmarked estimates were not statistically significant.}
\label{tab:case-study-perturbations}
\begin{tabular}{@{}
    >{\raggedright\arraybackslash}p{0.16\textwidth}
    >{\raggedright\arraybackslash}p{0.10\textwidth}
    >{\raggedright\arraybackslash}p{0.10\textwidth}
    >{\raggedright\arraybackslash}p{0.15\textwidth}
    >{\raggedright\arraybackslash}p{0.40\textwidth}
@{}}
\toprule
\textbf{Perturbation}
& \textbf{TRAILS}
& \shortstack[l]{\textbf{Context}\\\textbf{relevance}}
& \textbf{Outcome sensitivity}
& \textbf{Interpretation} \\
\midrule
\multicolumn{5}{@{}l}{\textbf{Repeated Prisoner's Dilemma}} \\
\midrule
Persona format
& TRAILS-R & Irrelevant
& $75.2^{***}$ pp
& Cooperation depended strongly on persona formatting. \\
\addlinespace[3pt]
Game framing
& TRAILS-R & Relevant
& $19.3^{***}$ pp
& Framing conditions should qualify the claim. \\
\addlinespace[3pt]
Memory representation
& TRAILS-R & Irrelevant
& $6.2^{***}$ pp
& Smaller but significant change; assess claim preservation against the tolerance. \\
\midrule
\multicolumn{5}{@{}l}{\textbf{Echo-chamber simulation}} \\
\midrule
Persona representation
& TRAILS-R & Irrelevant
& $|\Delta r|=0.194^{***}$\newline $|\Delta H|=0.051^{***}$
& Estimated collective structure depended on persona representation. \\
\addlinespace[3pt]
Activation probability
& TRAILS-D & Irrelevant
& $|\Delta r|=0.068^{***}$\newline $|\Delta H|=0.032^{***}$
& Significant changes in both outcomes; assess claim preservation against the tolerance. \\
\addlinespace[3pt]
Memory window
& TRAILS-D & Irrelevant
& $|\Delta r|=0.009$\newline $|\Delta H|=0.011$
& Small estimated differences; claim preservation requires a tolerance-based assessment. \\
\addlinespace[3pt]
Recommendation feed size
& TRAILS-D & Irrelevant
& $|\Delta r|=0.001$\newline $|\Delta H|=0.006$
& Small estimated differences; claim preservation requires a tolerance-based assessment. \\
\bottomrule
\end{tabular}
\end{table*}

\paragraph{Echo-chamber simulation.} We audited the claim that the simulation produced echo-chamber-like interaction structure among 100 agents over 15 rounds with fixed stances. Persona representation (\textsc{Bullet} versus \textsc{Prose}) was a context-irrelevant TRAILS-R perturbation. Under \textsc{Prose}, cross-group targeting increased by $5.15$ percentage points (95\% CI $[4.22,6.07]$), and cumulative stance assortativity was $0.194$ lower by round 15; the absolute difference in weighted same-group edge ratio was $0.051$. These results showed implementation dependence in the measured strength of echo-chamber structure. Activation probability, memory window, and recommendation feed size were treated as context-irrelevant TRAILS-D perturbations for this claim. Activation probability significantly changed both collective outcomes, whereas memory window and recommendation feed size produced smaller, non-significant differences. For the original phenomenon-presence claim, the verdict remained \textit{Inconclusive}: changes in outcome magnitude did not establish whether conditions crossed a specified criterion for echo-chamber presence.

\section{Simulation Prompts}
\label{appendix:prompts}

\subsection{Prisoner's Dilemma Prompts}

\subsubsection{Base Persona}

Across the Prisoner's Dilemma prompt-perturbation experiments reported in the paper, we use the following strategic persona:

\begin{promptbox}
You are a strategic participant focused on maximizing your long-term payoff. Study opponent behavior for exploitable patterns, and adjust your strategy to gain advantage while staying competitive.
\end{promptbox}

\subsubsection{Base Single-Agent Prompt Template}

In the single-agent setting, the model plays against a fixed external opponent policy. The opponent policy is not revealed to the model.

\begin{promptbox}
\textbf{Persona:}\\
\texttt{[BASE PERSONA]}\\
\vspace{0.8em}

\textbf{Game setup:}\\
- Repeated Prisoner's Dilemma for \texttt{[HORIZON]} rounds.\\
- Each round, both players simultaneously choose one action: \texttt{Cooperate} or \texttt{Defect}.\\
- You cannot observe the opponent's current-round action before choosing.\\
- Opponent follows a fixed external policy, unknown to you.\\
\vspace{0.8em}

\textbf{Payoffs:}\\
- If both choose \texttt{Cooperate}: \texttt{(3, 3)}.\\
- If you choose \texttt{Cooperate} and opponent chooses \texttt{Defect}: \texttt{(0, 5)}.\\
- If you choose \texttt{Defect} and opponent chooses \texttt{Cooperate}: \texttt{(5, 0)}.\\
- If both choose \texttt{Defect}: \texttt{(1, 1)}.\\
\vspace{0.8em}

\textbf{Task:}\\
Choose a reasonable action this round based on history and repeated-game context.\\
\vspace{0.8em}

\texttt{[HISTORY / MEMORY BLOCK]}\\
\vspace{0.8em}

\textbf{Output format:}\\
Return valid JSON with keys:\\
- \texttt{action}: one of \texttt{["Cooperate", "Defect"]}\\
- \texttt{reason}: short string
\end{promptbox}

\subsubsection{Base Two-Agent Prompt Template}

In the two-agent setting, both agents are model-driven players. Each agent receives the same prompt structure but with role-specific history.

\begin{promptbox}
\textbf{Persona:}\\
\texttt{[STRATEGIC PERSONA]}\\
\vspace{0.8em}

\textbf{Role:}\\
Your role label this run: Agent \texttt{[A/B]}.\\
\vspace{0.8em}

\textbf{Game setup:}\\
- Repeated Prisoner's Dilemma for \texttt{[HORIZON]} rounds.\\
- Each round, both agents simultaneously choose one action: \texttt{Cooperate} or \texttt{Defect}.\\
- You cannot observe the other agent's current-round action before choosing.\\
- The other agent is another model-driven player with the same prompt framing.\\
\vspace{0.8em}

\textbf{Payoffs:}\\
- If both choose \texttt{Cooperate}: \texttt{(3, 3)}.\\
- If you choose \texttt{Cooperate} and the other agent chooses \texttt{Defect}: \texttt{(0, 5)}.\\
- If you choose \texttt{Defect} and the other agent chooses \texttt{Cooperate}: \texttt{(5, 0)}.\\
- If both choose \texttt{Defect}: \texttt{(1, 1)}.\\
\vspace{0.8em}

\textbf{Task:}\\
Choose a reasonable action this round based on history and repeated-game context.\\
\vspace{0.8em}

\texttt{[ROLE-SPECIFIC HISTORY / MEMORY BLOCK]}\\
\vspace{0.8em}

\textbf{Output format:}\\
Return valid JSON with keys:\\
- \texttt{action}: one of \texttt{["Cooperate", "Defect"]}\\
- \texttt{reason}: short string
\end{promptbox}

\subsubsection{Output Schema and Parsing Instruction}

All Prisoner's Dilemma prompts end with the same JSON parsing instruction.

\begin{promptbox}
\textbf{Output format:}\\
Return valid JSON with keys:\\
- \texttt{action}: one of \texttt{["Cooperate", "Defect"]}\\
- \texttt{reason}: short string
\end{promptbox}

If a model response cannot be parsed as valid JSON, the parser attempts to extract a JSON object from the raw text. If parsing still fails, the fallback action is \texttt{Defect}.

\subsection{Echo-Chamber Prompts}

\subsubsection{Agent Persona Template}

Each echo-chamber agent is initialized with a minimal persona containing an agent ID, a short bio, a follower count, and a fixed stance toward AI regulation. Stance is represented on a five-point scale from strong opposition to strong support:

$$
s_i \in \{-2,-1,0,1,2\}.
$$

\begin{promptbox}
\texttt{\{}\\
\hspace*{1em}\texttt{"agent\_id": "[AGENT ID]",}\\
\hspace*{1em}\texttt{"bio": "[SHORT BIO]",}\\
\hspace*{1em}\texttt{"follower\_count": [FOLLOWER COUNT],}\\
\hspace*{1em}\texttt{"ai\_regulation\_stance": [-2|-1|0|1|2]}\\
\texttt{\}}
\end{promptbox}

The stance scale is:

\begin{promptbox}
\texttt{-2} = strongly oppose stronger AI regulation\\
\texttt{-1} = somewhat oppose stronger AI regulation\\
\texttt{0} = mixed or undecided\\
\texttt{1} = somewhat support stronger AI regulation\\
\texttt{2} = strongly support stronger AI regulation
\end{promptbox}

\subsubsection{Round Decision Prompt}

At each round, an activated agent receives its own profile, the number of neighbors in the fixed network, and a feed consisting only of recent posts from neighboring agents.

\begin{promptbox}
You are a social media user discussing one issue: whether governments should impose stronger regulation on advanced AI systems.\\
\vspace{0.8em}

\textbf{Your profile:}\\
- \texttt{agent\_id}: \texttt{[AGENT ID]}\\
- \texttt{bio}: \texttt{[BIO]}\\
- \texttt{follower\_count}: \texttt{[FOLLOWER COUNT]}\\
- current stance on AI regulation: \texttt{[STANCE]} where \texttt{-2} strongly oppose, \texttt{-1} somewhat oppose, \texttt{0} mixed or undecided, \texttt{1} somewhat support, \texttt{2} strongly support.\\
- neighbor count in the network: \texttt{[NEIGHBOR COUNT]}\\
\vspace{0.8em}

\textbf{Simulation rules:}\\
- This is a fixed network for this run.\\
- You only see recent posts from your neighbors.\\
- Keep behavior simple and realistic.\\
- Do not invent facts. Use short social-media-style language.\\
- You may choose exactly one action this round: \texttt{POST}, \texttt{REPOST}, \texttt{REPLY}, or \texttt{SILENT}.\\
- \texttt{REPOST} or \texttt{REPLY} should only target one visible post from the feed below.\\
- Your stance is fixed for the entire run and must not change.\\
- Set \texttt{updated\_stance} equal to your current stance.\\
\vspace{0.8em}

\textbf{Round:} \texttt{[ROUND]} / \texttt{[TOTAL ROUNDS]}\\
\vspace{0.8em}

\textbf{Visible feed from your neighbors:}\\
\texttt{[VISIBLE FEED]}\\
\vspace{0.8em}

\textbf{Return valid JSON with exactly these keys:}\\
\texttt{\{}\\
\hspace*{1em}\texttt{"action": "POST|REPOST|REPLY|SILENT",}\\
\hspace*{1em}\texttt{"target\_index": integer or null,}\\
\hspace*{1em}\texttt{"message": "short text, or empty string if SILENT",}\\
\hspace*{1em}\texttt{"updated\_stance": -2|-1|0|1|2,}\\
\hspace*{1em}\texttt{"reason": "brief reason"}\\
\texttt{\}}\\
\vspace{0.8em}

If \texttt{target\_index} is used, it must refer to one visible feed item by its 1-based index.
\end{promptbox}

\subsubsection{Visible Feed Format}

When neighbor posts are visible, they are rendered as a numbered list.

\begin{promptbox}
\texttt{[1] post\_id=[POST ID] | author=[AUTHOR ID] | action\_type=[POST/REPOST/REPLY] | text=[POST TEXT]}\\
\texttt{[2] post\_id=[POST ID] | author=[AUTHOR ID] | action\_type=[POST/REPOST/REPLY] | target\_post\_id=[TARGET POST ID] | text=[POST TEXT]}\\
\texttt{\ldots}
\end{promptbox}

If the agent has no visible recent neighbor posts, the feed is rendered as:

\begin{promptbox}
No recent posts from your neighbors.
\end{promptbox}

\subsubsection{Output Schema and Parsing Instruction}

All echo-chamber prompts require the following JSON schema.

\begin{promptbox}
\texttt{\{}\\
\hspace*{1em}\texttt{"action": "POST|REPOST|REPLY|SILENT",}\\
\hspace*{1em}\texttt{"target\_index": integer or null,}\\
\hspace*{1em}\texttt{"message": "short text, or empty string if SILENT",}\\
\hspace*{1em}\texttt{"updated\_stance": -2|-1|0|1|2,}\\
\hspace*{1em}\texttt{"reason": "brief reason"}\\
\texttt{\}}
\end{promptbox}

The parser accepts only four actions:

$$
a_i^t \in \{\texttt{POST}, \texttt{REPOST}, \texttt{REPLY}, \texttt{SILENT}\}.
$$

If \texttt{REPOST} or \texttt{REPLY} is selected, the target must refer to a visible feed item by its one-indexed position. The agent's stance is frozen throughout the run, so \texttt{updated\_stance} is constrained to equal the current stance:

$$
s_i^{t+1} = s_i^t.
$$

%%%%%%%%%%%%%%%%%%%%%%%%%%%%%%%%%%%%%%%%%%%%%%%%%%%%%%%%%%%%

% No checklist for position paper
% \newpage
% \input{checklist.tex}

\end{document}